\documentclass[a4paper, 10pt]{article}

\usepackage{epsfig,amssymb,euscript,xspace,color}
\usepackage{amsmath,jheppub,mathtools}
\def\bea{\begin{eqnarray}}
\def\eea{\end{eqnarray}}
\def\be{\begin{equation}}
\def\ee{\end{equation}}

\newcommand{\bra}{\langle}
\newcommand{\ket}{\rangle}
\newcommand{\rpl}{r_{\!+}}
\newcommand{\rmi}{r_{\!-}}
\newcommand{\Ll}{\mathcal{L}}
\newcommand{\Llo}{\mathcal{L}_{\varphi_1}}
\newcommand{\Llt}{\mathcal{L}_{\varphi_2}}


\def\cF{{\cal F}}

\def\cO{{\cal O}}

\def\cQ{{\cal Q}}
\def\cL{{\cal L}}

\def\cS{{\cal S}}

\def\cL{{\cal L}}
\def\cA{{\cal A}}
\def\cB{{\cal B}}

\def\cH{{\cal H}}

\def\cF{{\cal F}}
\def\cG{{\cal G}}

\begin{document}

\title{JT gravity and near-extremal thermodynamics for Kerr black holes in $AdS_{4,5}$ for rotating perturbations}
\author{ Rohan R. Poojary$^a$}
\affiliation{$^a$Institute for Theoretical Physics, TU Wien,\\
Wiedner Hauptstrasse 8-10, 1040 Vienna, Austria.  }
\emailAdd{rpglaznos@gmail.com}

\abstract{
We study the near horizon 2d gravity theory which captures the near extremal thermodynamics of Kerr black holes where a linear combination of excess angular momentum $\delta J $ and  excess mass $\delta M$ is held fixed. 
These correspond to processes where  both the mass and the angular momenta of  extremal Kerr black holes are perturbed leaving them near extremal.  For the Kerr $AdS_4$ we hold $\delta J-\Ll\,\delta M=0 $ while for Myers-Perry(MP) type Kerr black hole in  $AdS_5$ we hold $\delta J_{\varphi_{1,2}}\hspace{-0.2cm}-\Ll_{\varphi_{1,2}}\,\delta M=0$.   
We show that in near horizon, the 2d Jackiw-Teitelboim theory is able to capture the thermodynamics of the  higher dimensional black holes at small near extremal temperatures $T_H$. 
We show this by generalizing the near horizon limits found in literature by parameters $\Ll$ and $\Ll_{\varphi_{1,2}}$ for the two geometries. The resulting JT theory captures the near extremal thermodynamics of such  geometries provided we identify the temperature $T^{(2)}_H$ of the near horizon $AdS_2$ geometry  to be $T^{(2)}_H=T_H/(1-\mu\,\Ll)$ for 4d Kerr and $T^{(2)}_H=T_H/(1-\mu\,(\Llo+\Llt))$ for 5d Kerr where $\mu$  is their chemical potential, with $\mu\,\Ll<1$ and $\mu\,(\Llo+\Llt)<1$ respectively. We also argue that such a theory embeds itself non-trivially in the higher dimensional theory dual to the Kerr geometries. 
}

\maketitle


\section{Introduction}
Strongly interacting holographic systems have proved a very useful arena for understanding chaotic dynamics of large-$N$ theories and phenomena related to thermalization. These have in turn developed our understanding of quantum gravity dual to such strongly coupled systems in the bulk. More specifically the chaotic behaviour of large-$N$ strongly interacting theories has long since been shown to be mimicked by the scrambling behaviour of black holes in the dual $AdS$ to  in-fallen perturbation \cite{Shenker:2013pqa,Shenker:2013yza,Shenker:2014cwa}. It was famously shown  by Shenker \& Stanford that the finely tuned mutual information $I[A:B]$ contained in the thermo-field double (TFD) state dual to a static black hole in $AdS$ with temperature $T_H$ is perturbed due an in-fallen perturbation of $\cO(G_N)$ at a rate controlled by 
\be
\lambda_L=\frac{2\pi}{\beta},\hspace{0.3cm}\beta=T_H^{-1}.
\label{SS_Lyapunov}
\ee 
This mutual information is completely `scrambled' by a time scale known as the scrambling time $t_*$ given by
\be
t_*\sim\frac{\beta}{2\pi} \log G_N
\ee
Here $A$ \& $B$ are large enough subsystems belonging to the $left$ \& $right$ CFTs constituting the TFD. See \cite{Leichenauer:2014nxa} for a generalization of this computation to Reissner-Nordst\"{o}rm black holes in $AdS_4$. This phenomena of scrambling was also further analysed using 4pt \emph{out-of-time-ordered-correlators}(OTOCs) both in static BTZ black holes \cite{Shenker:2014cwa} and in the dual $CFT_2$ at finite temperature assuming  vacuum block domination  \cite{Roberts:2014ifa}. Here the scrambling behaviour is captured by the exponential behaviour of the OTOC as
\be
\frac{\bra V(t)W(0)V(t)W(0)\ket_{\rm OTO}}{\bra VV\ket\bra WW\ket}=1-\frac{\#}{c}e^{\lambda_L t}+\cO(c^{-2}),c\sim \ell^{d-2}/G_N
\label{OTOC_growth}
\ee
where $\#$ is a numerical factor depending on the details of the operators $V$ and $W$ in the CFT and $\ell$ being the radius of the $AdS_d$ dual. It was shown using assumptions of unitarity and boundedness for a QFT at a temperature $T_H$ that the scrambling time as seen by the OTOC is bounded by the fact that the Lyapunov index $\lambda_L$ is always bounded to be $\lambda_L<2\pi T_H$ \cite{Maldacena:2015waa} which came to be known as the MSS bound. This bound was shown to be saturated for the Nambu-Goto dynamics of a string stretched in BTZ \cite{deBoer:2017xdk} and for the case of critical quenches in CFT$_2$ with large central charge \citep{Das:2021qsd}. Explicit results for both the Lyapunov index and the butterfly velocity($v_B$)- the corresponding spread in the spatial direction, have been recently computed for a driven large-$c$ $CFT_2$ \cite{Das:2022jrr}. Here $\lambda_L$ and $v_B$ depend on the characteristics of the drive protocol. These have further been understood from a bulk perspective using $AdS_2$ brane probing an $AdS_3$ \cite{Das:2022pez}.
\\\\
The solvable 1d Sachdev-Yi-Kitaev(SYK) models have been shown to exhibit a similar chaotic behaviour for its OTOCs \cite{Maldacena:2016hyu}. This strongly interacting model has a 1d conformal symmetry at its zero temperature (IR) limit. At small temperatures this conformal symmetry is spontaneously broken giving rise to Goldstone modes parametrized by 1d diffeomorphisms.  These Goldstone modes were shown to be responsible for the chaotic behaviour of the OTOCs and are governed by a Schwarzian action along the Euclidean time circle. This low energy or `soft sector' of SYK models was then shown to be holographically dual to a 2d gravity theory first studied by Jackiw and Teitelboim and hence called the Jackiw-Teitelboim(JT) theory \cite{Jensen:2016pah,Maldacena:2016upp}\footnote{For the case of the Nambu-Goto string in BTZ the Schwarzian sector was analysed in \cite{Banerjee:2018twd,Banerjee:2018kwy}. The chaotic behaviour of   DBI brane dynamics in $AdS_5\times S^5$ were also worked out here. The $\widehat{\rm CGHS}$ model was also shown to be dual to the SKY model in the limit of large specific heat and vanishing compressibility \cite{Afshar:2019axx}. }. The JT theory is a theory of deformations of $AdS_2$ and consists of a metric and a dilaton\footnote{The JT theory is a special case of the well studied genric dilaton gravity models, these have been analysed both in gauge theoretic and metric formulations $c.f.$ \cite{Grumiller:2022qhx} for a review.}. The OTOC of the dual 1d CFT in such a theory was then shown to have a similar chaotic behaviour with $\lambda_L$ given by  \eqref{SS_Lyapunov} where the $AdS_2$ had a horizon given by
\be
ds^2_{2d}=\frac{dR^2}{(R^2-\delta\rpl^2)}+(R^2-\delta\rpl^2)dT^2,\hspace{0.4cm} \delta\rpl=2\pi T_H
\label{ds2_JT}
\ee   
It was subsequently shown that JT theory captures the dynamics and thermodynamics of higher dimensional near extremal black holes in both flat and $AdS$ spaces \cite{Banerjee:2019vff,Nayak:2018qej,Moitra:2018jqs,Moitra:2019bub,Castro:2018ffi}. 
Specifically the JT theory was shown to capture the thermodynamics of near extremal charged and rotating black holes where in an extremal black hole is perturbed by changing its mass while its extremal angular momentum and/or charge is held fixed \cite{Nayak:2018qej,Moitra:2019bub}\footnote{Readers can also refer to \cite{Johnstone:2013ioa} for thermodynamical analysis of extremal black hole and \citep{Hajian:2013lna} for near extremal black hole geometries. }. The JT gravity theory was shown to reproduce the excess mass and entropy as a fucntion of temperature close to extremality of the higher dimensional black hole. 
It turns out that the JT theory coupled to 2d matter can be solved exactly \cite{Yang:2018gdb,Kitaev:2018wpr} which is useful since the quantum effects associated with the gravitational degrees of freedom is this region are important \cite{Almheiri:2014cka,Jensen:2016pah,Maldacena:2016upp,Engelsoy:2016xyb}. This is known to modify the naive classical gravity results at low temperatures. The density of states vanishes continuously as the geometry approaches extremality for non-supersymmetric configurations \cite{Bagrets:2016cdf,Kitaev:2018wpr,Stanford:2017thb}. However for supersymmetric configurations,  there exists a  gap in the density of states and a large degeneracy of states at zero energy over extremality \cite{Boruch:2022tno}. This has recently lead to some interesting observations in supersymmetric extensions of the JT model \cite{Lin:2022zxd,Lin:2022rzw}. 
\\\\
The case for rotating black holes is a bit intriguing too. For the case of rotating BTZ its chaotic behaviour was first analysed in \cite{Poojary:2018esz,Stikonas:2018ane,Jahnke:2019gxr} where 2 possible Lyapunov indices were observed corresponding to left and right moving temperatures of the CFT$_2$, one of which is greater than the BTZ temperature and survives extremality. It was further shown in \cite{Halder:2019ric} that the arguments used to derive the MSS bound when generalized to the case of thermal QFT with a chemical potential $\mu$ corresponding to a charge associated with a global symmetry yields a modified bound on the Lyapunov index
\be
\lambda_L=\frac{2\pi T_H}{(1-\mu/\mu_c)}
\label{Halders_bound}
\ee
where $\mu_c$ corresponds to the maximum value attainable by $\mu$\footnote{For the case of holographic systems this would correspond to the chemical potential $\mu$ attained at extremality.}. 
It was later shown that long time behaviour of the OTOC for systems dual to rotating BTZ shows an average growth controlled  by $|\lambda_L|=2\pi T_H$ \cite{Mezei:2019dfv,Craps:2020ahu,Craps:2021bmz}, thus implying that the scrambling time as seen by the OTOC in the case of CFT$_2$ dual to rotating BTZ the scrmabling time is still controlled by the temperature of the black hole. The finely tuned mutual information $I[A:B]$ contained in the TFD state is a better diagnostic of late time perturbation as its value is known to be bounded by below by 4pt correlators \cite{Wolf:2007tdq}. It was recently shown \cite{Malvimat:2021itk} that for rotating in-falling perturbation  in a  rotating BTZ, $I[A:B]$ is perturbed at a rate governed by
\be
\lambda_L=\frac{2\pi T_H}{(1-\mu\,\Ll)}
\label{Lyapunov_rot_BTZ_L}
\ee 
Here $\Ll$ being the angular momentum per unit energy of the perturbation is always bounded by $\Ll<\mu^{-1}$. This generalizes the result of Shenker and Stanford \cite{Shenker:2013pqa} for the case of rotating shockwaves. This computation has since also been generalized for the case rotating equatorial shockwaves perturbing the mutual information in Kerr $AdS_4$ and Myers-Perry Kerr $AdS_5$ \cite{Malvimat:2022fhd,Malvimat:2022oue}. In the case of Kerr $AdS_4$  we find that at late times the minimum value of the instantaneous  $\lambda_L$ is bounded by \eqref{Lyapunov_rot_BTZ_L} but can clearly be greater than \eqref{SS_Lyapunov} for certain values of $\mu$. The long time average\footnote{Here by long time average we mean the time taken by the perturbation to completely cancel the value of $I[A:B]$ for the original geometry.} relevant for computing the scrambling time yields  value given by \eqref{Lyapunov_rot_BTZ_L}. The findings in the case of the Myers-Perry Kerr black holes in $AdS_5$ are similar to those for Kerr $AdS_4$ with $\Ll=\Llo+\Llt$; $\Llo$ \& $\Llt$ being the specific angular momenta of the perturbation along the symmetry directions $\varphi_1$ \& $\varphi_2$ of the Kerr black black hole in $AdS_5$.
\\\\
Given the above results for the observed value of $\lambda_L>2\pi T_H$ for Kerr black holes in $AdS$ a natural question arises with regards to a near horizon 2d gravity theory. Since the Lyapunov index is computed for late times leading upto the scrambling time of order $\log G_N$, it is expected that the IR of holographic theory plays a vital role. This corresponds to the region close to the horizon and in the extremal limit of black holes in $D$ dimensions this region is shown to have geometry of $AdS_2\times S^{(D-2)}$ \cite{Kunduri:2007vf,Figueras:2008qh}. The near extremal deformations of such near horizon geometries are captured by the JT theory as mentioned before. One can therefore ask can similar near horizon effective dynamics be ascribed to the observed value of $\lambda_L$ given by \eqref{Lyapunov_rot_BTZ_L} for Kerr black holes in $AdS_{4,5}$? In this paper we attempt to obtain such a near horizon effective prescription in terms of a 2d JT theory for Kerr $AdS_4$ and MP Kerr $AdS_5$. We restrict to dealing with the thermodynamics showing that such a near horizon prescription accurately captures the thermodynamics of near extremal black holes which are perturbed not only with additional mass $\Delta M$ but also additional angular momentum $\Delta J$ over its extremal value such that the resultant black hole is near extremal. This is a higher dimensional generalization of a similar result  for the case of near extremal  BTZ \cite{Poojary:2022meo}.
\\\\
This paper is organized as follows: we briefly review in the next subsection-\ref{Intro_JT} basic features  of the JT theory and relegate computation of its extremal entropy $\cS_{ext}$, ADM mass $\Delta M$ and excess entropy over extremality $\Delta\cS$ to an appendix-\ref{JT_appendix}. In section-\ref{Kerr_AdS_4} we deal with Kerr $AdS_4$ black hole and first generalize the near horizon limit for extremal geometries prevalent in literature in subsection-\ref{Kerr_AdS_4_NH_limits}. We then derive the JT action from the higher dimensional Einstein-Hilbert action with negative cosmological constant in subsection-\ref{Kerr_AdS_4_JT_der}. The Kerr $AdS_4$ thermodynamics is computed for the process perturbing with additional mass and angular momentum about extremality in subsection-\ref{Kerr_AdS_4_Thermodynamics}. We then reproduce this from the near horizon JT theory perspective in subsection-\ref{Kerr_AdS_4_NHNext_JT_analysis}. In section-\ref{MP_Kerr_AdS_5} we turn to MP Kerr $AdS_5$ black holes and study its perturbation about extremality for additional mass $\Delta M$ and angular momenta $\Delta J_{\varphi_1}$ \& $\Delta J_{\varphi_2}$ in a similar manner. We end with some discussions and conclusions in the last section.
\subsection{Review of the Jackiw-Teitelboim model}
\label{Intro_JT}
As mentioned before the JT theory is useful in capturing departures away from extremality for black holes which are solutions to the vacuum Einstein's equations.
The JT gravity theory is described by a Euclidean action given by
\bea
I_{(2d)}^{(E)}&=&I_{(0)}+I_{JT},\cr&&\cr
I_{(0)}&=&-\frac{\alpha^{-1}}{16\pi G_{(2)}}\left(\int \!d^2x\,\,\sqrt{g_{(2)}}\,\,R_{(2)}+2\int_\partial\!dx\sqrt{h_{(2)}}\,K_{(2)}\right)\cr&&\cr
I_{(JT)}&=&-\frac{1}{16\pi G_{(2)}}\left\lbrace\int\!d^2x\,\sqrt{g_{(2)}}\,\phi\left(R_{(2)}+2\right)+2\int_\partial\!dx\,\sqrt{h_{(2)}}\,\phi\,K_{(2)}\right\rbrace
\label{JT_action}
\eea
where $I_{(0)}$ is the topological term and the $I_{(JT)}$ is the dynamical part of the theory. Here $I_{(JT)}$ is also accompanied with a boundary term proportional to the boundary value of the dilaton $\phi$ so as to make $I_{(JT)}$ on -shell finite. 
This term can be determined by the general prescription for holographic renormalization prevalent in $AdS/CFT$ \cite{Henningson:1998ey}. We would be working with Dirichlet boundary conditions for all the fields at the boundary of $AdS_2$. For analyses pertaining to the other possible set of boundary condition see \cite{Grumiller:2017qao,Godet:2020xpk}.
The $eom$ for the metric and the dilaton respectively are 
\bea
&&R_{(2)}+2=0,\cr&&\cr
&&\nabla_{\mu}\nabla_{\nu}\phi-g_{(2)\mu\nu}\nabla^2\phi+ g_{(2)\mu\nu}\phi=4\pi G_{(2)}T_{(2)\mu\nu}=0
\label{eom_JT}
\eea
Here $T_{(2)\mu\nu}$ is the 2d matter stress tensor resulting from the matter fields in the 2d theory. Here we assume that the dilaton does not couple to the matter fields\footnote{This can be proved by expanding the higher dimensional minimally coupled matter action about the extremal solution to linear order in dilaton. } 
As the metric  here is fixed to be locally $AdS_2$ we can let its line line element take the value
\bea
ds^2_{(2d)}&=&\frac{dR^2}{(R^2-\delta\rpl^2)}+(R^2-\delta\rpl^2)dT^2,
\label{ads_2_JT}
\eea
where we denote the $AdS_2$ horizon to be at $\delta\rpl$. This also associates a non-zero temperature with the 2d geometry of 
\be 
T^{(2)}_H=\delta\rpl/(2\pi).
\label{delta_rpl_T_2_H}
\ee 
This non-vanishing temperature is the directly related to the fact that we are describing a near extremal geometry in higher dimensions. The dilaton $eom$ can also be exactly solved giving
\bea
\phi&=&R\,\phi_\partial+ \sqrt{R^2-\delta\rpl^2}\left(c_2e^{\delta\rpl T}+c_3e^{-\delta\rpl T}\right),\hspace{0.2cm}
\label{dilaton_sol_gen_JT}
\eea
where $\phi_\partial,c_1\,\&\,c_2$ are constants.
As we would be interested in the thermodynamics of the black holes solving the  vacuum Einstein's equations we would only deal with the stationary solution  of the dilaton. It is worth noting that the $SL(2,\mathbb{R})$ symmetry seen in the $AdS_2$ metric \eqref{ads_2_JT} is broken down to a $U(1)$ once one picks an on-shell value for the dilaton. This breaking of scale invariance  can be seen as an effect of being at a small but finite temperature and we return to this shortly in discussing the extent of the near horizon throat region in the bulk of the higher dimensional black hole.
\\\\
In order for the JT theory to reproduce the dynamics of the higher dimensional near extremal black hole we need to furnish it with 2 pieces of information of the latter: 1) The near extremal temperature $T_H$ and 2) The size of the black hole $i.e.$ its outer horizon's radius.
To the leading order the second is basically the information contained in $G_{(2)}$ which depends on the extremal horizon volume and the higher dimensional Newton's constant $G_{N}$.  Given just this information we can find that the topological term reproduces the extremal entropy of the higher dimensional geometry
\be
I_{(0)}\overset{(OS)}{=}-\cS_{ext},
\label{Topo_S_ext_JT}
\ee
The first information relates the temperature of the higher dimensional near extremal black hole $T_H$ to $T^{(2)}_H$. For the cases studied till now in literature this was basically $T_H=T^{(2)}_H$. Given this information and how the dilaton $\phi$ relates to the horizon volume $V_H$ of the full higher dimensional geometry we get    
\be
\frac{\Delta\cS}{\cS_{ext}}=\alpha\,\phi|_{R=\delta\rpl}
\label{Delta_S_by_S_JT}
\ee
where $\alpha$ is a number that depends on the relation between $\phi$ and $V_H$ about extremality.
$\Delta\cS$ is the change in the entropy of the higher dimensional black hole over extremality and this depends on  $T_H$.
Similarly we can also reproduce the 2d ADM Mass given these information
\be
M^{(2)}= \frac{\delta\rpl}{16\pi G_{(2)}}\phi|_{R=\delta\rpl}=\Delta M
\label{Delta_M}
\ee
Here $\Delta M$ is the excess mass over extremality of the higher dimensional black hole and this too depends on $T_H$. 
Here $G_{(2)}$ captures the size of extremal black hole while $\phi|_{\delta\rpl}$ captures the fluctuation of this size over extremality.
Given how $G_{(2)},\phi|_{\delta\rpl}$ \& $T^{(2)}_H$ relate to the parent higher dimensional near extremal black hole, the JT theory correctly captures the temperature dependence of $\Delta\cS$ and $\Delta M$ and consequently the first law of black hole thermodynamics relevant to the higher dimensional black hole. This check has been done for a wide variety of near extremal geometries in BTZ\cite{Banerjee:2019vff,Ghosh:2019rcj}, RN black holes in asymptotically  flat and $AdS$ spaces \cite{Nayak:2018qej} Kerr black holes in flat and $AdS$ space in both 4 and 5 dim. \cite{Moitra:2019bub,Castro:2018ffi}.    
\\\\
It is worth noticing the extent of the near horizon throat region described by the $AdS_2$ geometry \eqref{ads_2_JT,Nayak:2018qej,Moitra:2019bub}. At strict extremality this region is infinitely long and stretches all the way till the conformal boundary of the higher dimensional $AdS$ black hole being analysed. 
The presence of the dilaton signals the breaking of the scale invariance present in the near horizon $AdS_2$.
At small temperatures the extent of the near horizon $AdS_2$ is determined by the finite temperature corrections at large $R$ as seen in the metric \eqref{ads_2_JT} and the growing on-shell value of the dilaton $\phi$ as seen for stationary configurations \eqref{dilaton_sol_gen_JT}. 
The former corrects the boundary metric of $AdS_2$ as $\sim(\beta^{(2)}R)^{-2}$ while the latter grows as $R\,\phi_\partial$. 
The location of the throat boundary is the region far from the outer horizon such that the corrections from the finite temperature have died down but not so far that the breaking of scale invariance due to the growth of the dilaton has become significant. We thus have the throat boundary determined by
\be
R\,\phi_\partial \ll 1 \ll \beta^{(2)}R 
\label{throat_boundary_limits}
\ee
It is worth pointing out again that the how the actual values of $\phi_\partial$ and $\beta^{(2)}$ relate to that of the near extremal higher dimensional black hole needs to be supplied by geometric arguments.
\\\\
The chaotic dynamics associated with probe scalar fields interacting with each other $via$ gravitational interactions for near extremal black holes is also captured by JT gravity. 
As mentioned in the introduction low dimensional systems like CFT$_1$ with Schwarzian dynamics, the wide variety of SYK-like models, holographic CFTs in 2d with vacuum domination exhibit a tell-tale sign of chaotic behaviour in terms of the exponential growth as seen in the OTOC \eqref{OTOC_growth}. 
Here $G_N\sim1/c$ with $c$ being the central charge associated with the CFT. 
In the context of holography this has only ever been explicitly checked for $AdS_3/CFT_2$ $i.e.$ BTZ black holes and $AdS_2/CFT_1$. 
In the case of higher dimensional black holes like Reissner-Nordstr\"{o}m, Kerr \& Kerr-Newman $AdS$ black holes in bulk dim$>3$, such a chaotic behaviour is also expected but explicit computation of OTOCs in the dual CFT$_{d-1}$ has proved intractable as the \emph{bulk-to-boundary} correlators of the bulk scalar fields is  difficult to compute. 
The apparent simplification in near extremal dynamics of such systems in terms of the JT model allows one to examine the late time behaviour of OTOCs. 
Here the OTOCs of the dual CFT$_{(d-1)}$ are mapped to the OTOCs of the $AdS_2/CFT_1$ of the JT model. 
As the latter describes the dynamics in the near horizon region, this corresponds to describing the low energy sector of OTOC in the full $AdS_d/CFT_{(d-1)}$. 
\\\\
It can be shown while obtaining the JT theory from higher dimensional the Einstein-Hilbert action about near extremal geometries that the probe scalar fields in the resulting 2d theory couple only 2d metric and not to dilaton $\phi$ of the JT theory\footnote{\emph{Note} this simplification happens only to linear order in dilaton interaction which is also the limit in which we obtain the JT theory.}. Therefore for some 2d scalar field $\chi_i$ we have
\be
I_{2d}^{matter}[\chi_i]=\int d^2x\sqrt{g_{(2)}}\,(\nabla\chi_i)^2+I_{count}[\chi_i]
\label{matter_JT}
\ee
where we have added holographic counter term to make the euclidean action on-shell finite.
Here the JT theory interacts with the scalar field in a $AdS_2$ background metric \eqref{ads_2_JT} $via$ the dilaton $eom$
\be
\nabla_{\mu}\nabla_{\nu}\phi-g_{(2)\mu\nu}\nabla^2\phi+ g_{(2)\mu\nu}\phi=4\pi G_{(2)}T_{(2)\mu\nu}(\chi_i)
\ee
$T_{\mu\nu}(\chi_i)$ being the stress-tensor associated with the scalar field $\chi_i$. 
For two such scalar fields $\chi_1$ \& $\chi_2$ interacting $via$ the above equation with each other  it was shown \cite{Jensen:2016pah,Maldacena:2016upp} that the effective action in the dual CFT$_1$ is given by the Schwarzian action which in turn governs the interaction between (2pt. function of) the dual operators $V$ \& $W$. Here  $V$ \& $W$ are the dual operators of bulk fields $\chi_1$ \& $\chi_2$ respectively.  The 1d Euclidean effective action is
\be
I_{eff}=-\frac{1}{16\pi G_{(2)}}\int_0^{\beta^{(2)}} \hspace{-0.5cm}dT\, \left[2\{f,T\}+\left(\frac{2\pi}{\beta^{(2)}}\right)^2f'^2\right]
\label{JT_Schwarzian}
\ee
The above action is understood as arising from the phase-space of solutions to $R_{(2)}+2=0$ about the $AdS_2$ metric \eqref{ads_2_JT} with temperature $T^{(2)}=\beta^{(2)-1}$ for Euclidean time reparametrizations $T\rightarrow f(T)$. 
For small deformations $T\rightarrow T+\epsilon(T)$ we can find the corrections to the 4pt function of dual operators $V$ \& $W$ given that \eqref{JT_Schwarzian} provides a action for the conformal transformation of the 2pt function under $T\rightarrow T+\epsilon(T)$.
\bea
&&\bra V(f_1) V(f_2)\ket=\left(\frac{\partial f(T_1)}{\partial T_1}\frac{\partial f(T_2)}{\partial T_2}\right)^{h_V/2}\bra V(T_1) V(T_2)\ket \hspace{0.3cm}{\rm for}\hspace{0.3cm} T\rightarrow T+\epsilon(T)\cr&&\cr
\implies&&\frac{\delta\bra V(T_1)V(T_2)\ket}{\bra V(T_1)V(T_2)\ket}=\cB^{(1)}_{h_V}(\epsilon_1,\epsilon_2)=h_V\left[\epsilon'(T_1)+\epsilon'(T_2)-\frac{\epsilon(T_1)-\epsilon(T_2)}{\tan(\pi T^{(2)}_H(T_1-T_2))}\right]
\eea
\be
\frac{\bra V(T_1)V(T_3)W(T_2)W(T_4)\ket}{\bra V(T_1)V(T_3)\ket\bra W(T_2)W(T_4)\ket}=1+\bra\cB^{(1)}_{h_V}(\epsilon_1,\epsilon_3)\cB^{(1)}_{h_w}(\epsilon_2,\epsilon_4)\ket
\ee
where we use the quadratic action for $\epsilon$s obtained by expanding \eqref{JT_Schwarzian} to quadratic order in $\epsilon$s  to get the propagator $\bra\epsilon(T_1)\epsilon(T_2)\ket$. The $\bra\epsilon\epsilon\ket$ propagator is Euclidean and careful use of the "$i-\epsilon$"  prescription for analytical continuation to Lorentzian times gives the OTOC behaviour as in \eqref{OTOC_growth} with 
\be
\lambda_L=\frac{2\pi}{\beta^{(2)}}
\label{JT_lambda_L}
\ee 
For the analysis prevalent in literature of the JT model arising out of near extremal limits of RN, Kerr \& Kerr-Newman black holes we have $T^{(2)}_H=T_H$. 
Thus it is in the above manner that the simplification afforded by the JT model captures the OTOC behaviour \eqref{OTOC_growth} in the  IR sector for higher dimensional black holes. 
\section{Kerr-$AdS_4$ }
\label{Kerr_AdS_4}
The Kerr metric in $AdS_4$ in the Boyer-Lindquist coordinates  \cite{Plebanski:1976gy,Carter:1968ks,Iyer:1994ys} takes the form
\bea
&&ds^2=\rho^2\left(\frac{dr^2}{\Delta}+\frac{d\theta^2}{\Delta_\theta}\right)-\frac{\Delta}{\rho^2}\left(dt-\frac{a\sin^2\theta}{\Xi}d\varphi\right)^2+\frac{\Delta_\theta\sin^2\theta}{\rho^2}\left(adt-\frac{r^2+a^2}{\Xi}d\varphi\right)^2\cr&&\cr
{\rm where}&&
\hspace{2cm}
\Delta=(r^2+a^2)(1+r^2/l^2)-2mr
,\hspace{0.3cm}
\rho^2=r^2+a^2\cos^2\theta\cr&&\cr
&&\hspace{2cm}\Delta_\theta=1-\frac{a^2}{l^2}\cos^2\theta,\hspace{2.5cm}\Xi=1-\frac{a^2}{l^2}.
\label{Kerr_Boyer_Lindquist_4}
\eea
The outer and inner horizons exist as the two real roots of $\Delta=0$.
A peculiarity of these coordinates is that the boundary coordinates at $r\rightarrow\infty$  rotate with an angular velocity of
\be
\Omega_\infty=-\frac{a}{l^2}
\label{vel_boundary_Kerr_4}
\ee
while the horizon angular velocity is
\be
\Omega_\phi=\frac{a\,\Xi}{\rpl^2+a^2}
\label{vel_horizon_Kerr_4}
\ee
The chemical potential associated with the thermodynamics of the black hole is given by the horizon velocity as measured by a stationary boundary observer \cite{Papadimitriou:2005ii} 
\be
\Omega_+=\Omega_{\phi}-\Omega_\infty=\frac{a(1+\rpl^2/l^2)}{\rpl^2+a^2}=\mu
\label{vel_actual_Kerr_4}
\ee
The temperature and entropy can be given in terms of $\{a,\rpl\}$ as
\be
2\pi T_H=\frac{\rpl^2-a^2+\rpl^2/\l^{-2}(3\rpl^2+a^2)}{2\rpl(\rpl^2+a^2)},\hspace{0.2cm}\mathcal{S}=\frac{\pi(\rpl^2+a^2)}{G_{4}\Xi}
\label{T_H_S_Kerr_4}
\ee

\subsection{Near Horizon limits}
\label{Kerr_AdS_4_NH_limits}
In this subsection we would derive the near horizon limits of near extremal Kerr geometry as seen by the an in-falling null perturbation with an angular momentum in along the axi-symmetric directions of Kerr $AdS_4$. The expectation being that since the rotating null shockwaves in Kerr $AdS_4$ produce the kind of scrambling as seen in \cite{Malvimat:2022oue}, the relevant near horizon physics would show up in such a limit.
\\\\
We begin with first constructing in-out going null geodesics $\xi^\mu\partial_\mu$ parametrized by constants 
\be
\xi^2=0,\hspace{1cm}g_{\mu\nu}\xi^\mu\zeta^\nu_t=\mathcal{E},\hspace{1cm}g_{\mu\nu}\xi^\mu\zeta^\nu_\varphi=\mathcal{L},\hspace{1cm}\xi^\mu K_{\mu\nu}\xi^\nu=\mathcal{Q}
\label{geodesic_cond_Kerr_4}
\ee
where $\zeta_t=\partial_t-\tfrac{a}{l^2}\partial_\varphi$ and $\zeta_\varphi=\partial_\varphi$ generate time translations  and axi-symmetric rotations respectively with respect to a stationary boundary observer.
Here $K_{\mu\nu}$ is the symmetric tensor constructed from the Killing-Yano tensor $f_{\mu\nu}=-f_{\nu\mu}$ which is a higher form symmetry present in Kerr geometries. $K_{\mu\nu}$ satisfies 
\be
\nabla^\mu K_{\mu\nu}=0,\hspace{0.3cm} K_{\mu\nu}=-f_{\mu\rho}\,f^\rho_{\,\,\,\nu}
\label{Kdd_Kerr_4}
\ee
We would want the geodesic vector fields to be an in-out going pair, which we label $\xi^\pm$(or $\xi_\pm$). The two are related by reversing the signs of $\mathcal{E}$ and $\Ll$ and $\xi^\theta$. We also choose to normalize $\mathcal{E}=1$ therefore $\Ll$ is the angular momentum per unit energy $i.e.$ specific angular momentum of the geodesies. Further we would like the line elements along the geodesies to be exact
\be
\xi^{-}_\mu dx^\mu = dv,\,\,\, \xi^{+}_\mu dx^\mu = du, \,\,\,\forall\, \theta\in [0,\pi]
\label{LC_coord_Kerr_4}
\ee
The $\{u,v\}$ coordinates are the light-cone coordinates with angular momentum $\Ll$.
For this it is enough to show
\be
\partial_\theta\xi^\pm_r=\partial_r\xi^\pm_\theta
\label{LC_coord_exactness_cond_Kerr_4}
\ee
This implies $\mathcal{Q}=const$\footnote{In this case $\partial_\theta\xi^\pm_r=\partial_r\xi^\pm_\theta=0$}.
A simple choice for $\cQ$ is the one in which equatorial geodesies  remain equatorial for all times $i.e.$ $\xi^\theta=0$ for such geodesies. This implies choosing
\be
\mathcal{Q}=-\frac{(a-\mathcal{L})^2}{a^2}
\label{Carters_Q_equatorial_Kerr_4}
\ee 
Using the above geodesies we can recast the metric line element \eqref{Kerr_Boyer_Lindquist_4} as
\bea
&&ds^2=F\,\xi_\mu^+\xi_\nu^- dx^\mu dx^\nu+ h\,(dz+h_\tau d\tau)^2 +g\, (d\theta+g_\tau d\tau)^2,\hspace{0.4cm} g_\tau(\pi/2)=0 
\label{Kerr_LC_4}\cr&&\cr
{\rm where} &&\hspace{0.3cm}\tau=t\left(1-\tfrac{a\mathcal{L}}{l^2}\right)-\mathcal{L}\,\varphi,\hspace{1.4cm} \eta \,z=(1+\mathcal{L}\gamma)\,\varphi-(1-\tfrac{a\mathcal{L}}{l^2})\gamma \,t\cr&&\cr
{\rm with}&&\hspace{0.2cm}\gamma=\frac{a(1-a^2/l^2)}{(a^2+\rpl^2)l^2-(\rpl^2+l^2)a\mathcal{L}},\hspace{0.3cm}\eta=\frac{l^2(1+\mathcal{L}\gamma)}{l^2-a\mathcal{L}}
\eea
Here the $\tau$ coordinate naturally occurs as the line element along the $\{t,\varphi\}$ directions in the one-form along null geodesic $i.e.$ $2d\tau=du-dv$. 
The above relation can be inverted to
\be
t=\frac{\tau+\Ll\, z}{1-\mu \,\mathcal{L}},\hspace{0.3cm}\varphi=\frac{(1-a\mathcal{L}/\l^2)\,z+\Omega_{\varphi}\tau}{1-\mu \,\mathcal{L}}.
\label{t_phi_to_tau_z_Kerr_4}
\ee
We have defined the $z$ coordinate such that $h_\tau\sim\cO(r-\rpl)$ at the outer horizon and integrating the transverse $S^2$ defined by $\{\theta,z\}$ coordinates  with $\theta\in[0,\pi]$ and $z\in[0,2\pi]$ at $r=\rpl$ gives the horizon area. 
The tortoise coordinates are defined by integrating the $dr$ component of the one-form dual to the geodesies
\bea
&&du=\xi^+_\mu dx^\mu=dr_*-d\tau-\tilde{g}'d\theta \implies u=r_*-\tau-\tilde{g}(\theta)\cr&&\cr
&&dv=\xi^-_\mu dx^\mu=dr_*+d\tau+\tilde{g}'d\theta \implies v=r_*+\tau+\tilde{g}(\theta)\cr&&\cr
&&{\rm where}\hspace{0.3cm}\tilde{g}(\theta)=\frac{\sqrt{-\Xi\cot^2\theta(\ell^2(\Xi-\Delta_\theta)+\Ll^2\Delta_\theta)}}{\Delta_\theta},\hspace{0.3cm}\tilde{g}(\theta=\pi/2)=0.
\label{LC_coordinates_Kerr_4}
\eea
\be
\& \,\,\,\,\,dr_*=\tfrac{\sqrt{r \left(r^3 \left(\ell ^2-a^2\right) (\ell^2 -\mathcal{L}^2) +r \ell ^2 \left(\ell ^2-a^2\right) (a^2-\mathcal{L}^2) +2 M \ell ^4 (a-\mathcal{L})^2\right)}}{a^2 \left(r^2+\ell^2\right)+r \ell ^2 (r-2 M)+r^4}dr
\ee
The above light-cone $\{u,v\}$ coordinates are smooth at the horizon the vector field $\chi^\mu\partial_\mu$ that generates translation along $u$ (or $v$) fails to be affine at the horizon  
\bea
&&\chi\cdot\nabla\chi^\mu=\mathcal{K}\chi^\mu\cr&&\cr
{\rm where}&&\mathcal{K}=\tfrac{1}{2}\xi^+\cdot\partial F\hspace{0.5cm}{\rm with}\hspace{0.3cm}\kappa=\mathcal{K}\vert_{r=\rpl}\neq 0
\label{non_affine_Kerr_4}
\eea
One can therefore define affine coordinates at the outer horizon as
\be
U=-e^{\kappa(r_*-\tau-\tilde{g})},\,\,V=e^{\kappa(r_*+\tau+\tilde{g})}
\label{Kruskal_coordinates_equator_Kerr_4}
\ee
where the value of $\kappa$ is
\bea
&&\kappa=\frac{\kappa_0}{(1-\mu\,\Ll)},\,\,\,\,{\rm where}\,\,\mu=\Omega_+\,\,\,\,\,\&\,\,\,\,\,\kappa\vert_{\Ll\rightarrow0}=\kappa_0=\frac{2\pi}{\beta}=2\pi T_H
\label{kappa_Kerr_4}
\eea
Here we denote $\kappa_0$ to be $2\pi$ times the black hole temperature $T_H$. By construction the vector fields generating translation along the $\{U,V\}$ coordinates are affine (or smooth) across the outer horizon, we therefore identify them as the Kruskal coordinates and the metric line element takes the form
\bea
&&ds^2=\frac{F}{\kappa^2 UV}dUdV + h\,(dz+h_\tau d\tau)^2 +g\, (d\theta+g_\tau d\tau)^2,\hspace{0.4cm} g_\tau(\pi/2)=0 \cr&&\cr
{\rm where}&&d\tau=\frac{1}{2UV}(UdV-VdU)-\tilde{g}'d\theta
\label{Kerr_Kruskal_4}
\eea
\emph{Note,} as the future (past) horizon is defined at $U=0$ (V=0), the line element $h_\tau d\tau$ does not blow up at the horizon as the $z$ coordinate had been specifically chosen so that $h_\tau\sim\cO(r-\rpl)\sim \cO(UV)$ at the outer horizon\footnote{$g_\tau\sim\cO(r-\rpl)$ at the outer horizon, hence $\theta$ was not in need of redefinition. }.
\\\\
The above form of the metric is well suited to study the Dray-'tHooft solutions which account for the back reaction caused due to an in-falling null shockwave at very late time \cite{Malvimat:2022oue}. Since we would be interested in taking the near horizon near extremal limits we choose to work with the $\{r,\tau,\theta,z\}$ coordinates using the relation \eqref{t_phi_to_tau_z_Kerr_4}.  In these coordinates we obtain the strict extremal limit by scaling  the $\{r,\tau\}$ coordinates relative to the black hole horizons $r_\pm$ as
\bea
&&\rmi=r_0-\lambda\, \mathcal{A}\,\delta\rpl,\hspace{0.3cm}\rpl=r_0+\lambda\, \mathcal{A}\,\delta\rpl,\hspace{0.3cm} r=r_0+\lambda\,\mathcal{A}\,R,\hspace{0.3cm}\tau=\frac{T}{\lambda}\cr&&\cr
&&{\rm where}\hspace{0.3cm}\mathcal{A}=\frac{2r_0^2\ell^2 \left(r_0^2+\ell ^2\right) \left(1-\mu_0\,\Ll\right)}{6 r_0^2 \ell ^2-3 r_0^4+\ell ^4}\hspace{0.2cm}\&\,\,\mu_0=\frac{\sqrt{(\ell^2-r_0^2)(\ell^2+3r_0^2)}}{2r_0\ell}
\label{strict_limits_ads_4}
\eea
here $\mu_0$ refers to the extremal value of the chemical potential $\mu$. The strict extremal limit is obtained by taking $\lambda\rightarrow 0$
\bea
ds^2_{AdS_4,NHext}&=&\cF_{\!\!_{r_0}}(\theta)\left[\frac{dR^2}{R^2}-R^2dT^2\right]+\mathcal{G}_{\!_{r_0}}(\theta)\,d\theta^2+\frac{\sin^2\theta}{\cG_{\!_{r_0}}(\theta)}\left(dz+A_{_{r_0}T}\,dT\right)^2
\label{ads_4_ext}\cr&&\cr
{\rm where}&&\mathcal{F}_{\!\!_{r_0}}(\theta)=\frac{r_0^2 \ell ^2 \left(r_0^2+3 \ell ^2+ \left(3 r_0^2+\ell ^2\right)\cos (2 \theta )\right)}{12 r_0^2 \ell ^2+2 \ell ^4-6 r_0^4}\cr&&\cr
&&\mathcal{G}_{\!_{r_0}}(\theta)=\ell ^2-\frac{2\ell^2( \ell^4 -r_0^4)}{3 r_0^4+3 r_0^2 \ell ^2-2 \ell ^4+r_0^2  \left(3 r_0^2+\ell ^2\right)\cos (2 \theta )}\cr&&\cr
&&A_{_{r_0}T}=\frac{ \left(\ell^2-r_0^2\right) \left(3 r_0^2-\ell ^2\right) \sqrt{-3 r_0^4+2 r_0^2 \ell ^2+\ell ^4}}{3 r_0^6-9 r_0^4 \ell ^2+5 r_0^2 \ell ^4+\ell
   ^6}R
\eea
This is precisely the near horizon metric of extremal Kerr $AdS_4$ obtained in literature \cite{Lu:2008jk}. It is also worth pointing out that the free parameter $\Ll$ which dictates the relation between the Boyer-Lindquist coordinates $\{t,\varphi\}$ and $\{\tau,z\}$ as in \eqref{t_phi_to_tau_z_Kerr_4} does not make an explicit appearance. It is also worth emphasising that the $z$ coordinate needs no redefinition before scaling as it is already the smooth coordinate at the near horizon region. For $\Ll=0$ case $z$ is the near horizon angular coordinate found in literature. Therefore  we have obtained a near horizon limit wherein we approach the outer horizon along in going null geodesies with  a non zero specific angular momentum $\Ll$, and the cases studied till now in literature correspond to the setting $\Ll=0$. It must be pointed out that not any value of $\Ll$ is permitted as not all null rotating geodesies can reach the outer horizon having emenated from the $AdS_4$ boundary. The allowed values of $\Ll$ have to be determined by turning point analysis however it suffices for us to note that $\Ll$ is bounded by 
\be
\Ll<\mu^{-1}
\label{L_bound_kerr_4}
\ee   
for any Kerr geometry in $AdS_4$. In taking the above limit \eqref{strict_limits_ads_4} we have made sure that  there are no relative constant factors between $\tau$ and $T$. This is necessary as we would like to associate the same temperature to the near horizon $T$ coordinate as seen by the $\tau$ coordinate when both are Euclideanised. 
If one were to Eulcideanise the line element \eqref{Kerr_LC_4} in $\{r,\tau,\theta,z\}$ coordinates by substituting $\tau\rightarrow i\tau_{_{\!E}}$ then demanding smoothness at the outer horizon would imply that $\tau_{_{\!E}}$ would have periodicity of $\left(\frac{\kappa}{2\pi}\right)^{-1}$ where $\kappa$ is given in \eqref{kappa_Kerr_4}. This should not be surprising as demanding smoothness or `affineness' in the Lorentzian metric must imply smoothness or lack of a conical deficit in the Euclidean metric. 
\\\\
We can also obtain a simultaneous near horizon and near extremal limit by choosing to scale the $\{r,\tau\}$ coordinates as
\bea
&&\rmi=\rpl-2\lambda\, \mathcal{A}\,\delta\rpl,\hspace{0.3cm} r=\rpl+\lambda\,\mathcal{A}(R-\delta\rpl),\hspace{0.3cm}\tau=\frac{T}{\lambda}\cr&&\cr
&&{\rm where}\hspace{0.3cm}\mathcal{A}=\frac{2\rpl^2\ell^2 \left(\rpl^2+\ell ^2\right) \left(1-\mu\,\Ll\right)}{6 \rpl^2 \ell ^2-3 r_+^4+\ell ^4}
\label{limits_ads_4}
\eea
Note here $\cA$ is a function of the near extremal outer horizon $\rpl$. We get the near extremal  near horizon metric as 
\bea
ds^2_{AdS_4,NHNext}&=&\mathcal{F}_{\!\!_{\rpl}}(\theta)\left[\frac{dR^2}{(R^2-\delta\rpl^2)}-(R^2-\delta\rpl^2)dT^2\right]+\mathcal{G}_{\!_{\rpl}}(\theta)\,d\theta^2+\frac{\sin^2\theta}{\cG_{\!_{\rpl}}(\theta)}\left(dz+A_{_{\rpl}T}\,dT\right)^2\cr&&\cr
{\rm where}&&\mathcal{F}_{\!\!_{\rpl}}(\theta)=\frac{\rpl^2 \ell ^2 \left(\rpl^2+3 \ell ^2+ \left(3 \rpl^2+\ell ^2\right)\cos (2 \theta )\right)}{12 \rpl^2 \ell ^2+2 \ell ^4-6 \rpl^4}\cr&&\cr
&&\mathcal{G}_{\!_{\rpl}}(\theta)=\ell ^2-\frac{2\ell^2( \ell^4 -\rpl^4)}{3 \rpl^4+3 \rpl^2 \ell ^2-2 \ell ^4+\rpl^2  \left(3 \rpl^2+\ell ^2\right)\cos (2 \theta )}\cr&&\cr
&&A_{_{\rpl}T}=\frac{ \left(\ell^2-\rpl^2\right) \left(3 \rpl^2-\ell ^2\right) \sqrt{-3 \rpl^4+2 \rpl^2 \ell ^2+\ell ^4}}{3 \rpl^6-9 \rpl^4 \ell ^2+5 \rpl^2 \ell ^4+\ell
   ^6}(R-\delta\rpl)
\label{ads_4_next}
\eea
Here The 2d horizon at $R=\delta\rpl$ indicates the fact that original black hole is near extremal.
One could further expand the functions $\cF_{\!\rpl},\cG_{\rpl},A_{\rpl,T}$ in $\lambda$ as $\rpl\rightarrow r_0+\lambda \cA \delta\rpl$ where $\cA$ takes the value in \eqref{strict_limits_ads_4} in which case we can replace $\rpl\rightarrow r_0$ as $\lambda\rightarrow 0$. This would then correspond to the near horizon metric obtained in \cite{Lu:2008jk} for near extremal geometries for the case of $\Ll=0$. 
\\\\
One can define near horizon Kruskal coordinates $\{U^{(2)},V^{(2)}\}$ as
\be
dU^{(2)}=\frac{dR}{R^2-\delta\rpl^2}-dT,\hspace{0.3cm}dV^{(2)}=\frac{dR}{R^2-\delta\rpl^2}+dT
\label{NH_ads_2_Kruskal_kerr_4}
\ee
and relate them to $\{U,V\}$ obtained earlier in \eqref{Kruskal_coordinates_equator_Kerr_4}.
Using the limit \eqref{limits_ads_4} we see that the regions close to the outer horizon $U=0=V$ gets mapped to the entire domain of the $\{U^{(2)},V^{(2)}\}$ as $\lambda\rightarrow 0$
\be
U\rightarrow (U^{(2)})^{1/\lambda},\hspace{0.3cm}V\rightarrow (V^{(2)})^{1/\lambda}
\label{NH_ads_2_Kruskal_BL_Kruskal_Kerr_4}
\ee
Therefore we the above near horizon limits are basically correspond to approaching the outer horizon along null rotating geodesies with specific angular momentum $\Ll$ and one obtains limits prevalent in literature for the $\Ll=0$ case. Further the smoothness or `affineness' of the $\{U,V\}$ coordinates at $\rpl$ implies that the 2d temperature $T^{(2)}_H$ is equal to the temperature seen by the null geodesic $U=0$ (or $V=0$) $i.e.$ $\kappa/(2\pi)$. Therefore we have
\be
T_H^{(2)}=\frac{T_H}{1-\mu_0\,\Ll}=\frac{\delta\rpl}{2\pi}
\label{T_2_T_H_Kerr_4}
\ee
where $\mu_0$ is the extremal value of the chemical potential $\mu$. It is worth emphasising that any relation  between the near horizon temperature to that seen at the stationary boundary of asymptotic $AdS_4$ would require the knowledge of how the near horizon limit was taken. As we approached the outer horizon along a specific geodesic the relevant form of the metric is \eqref{Kerr_Kruskal_4}. Indeed it is the $dUdV$ component of the metric in \eqref{Kerr_Kruskal_4} that essentially becomes the $\{R,T\}$ component of the near horizon metric \eqref{ads_4_ext} \& \eqref{ads_4_next}. Therefore the above expression for $T_H^{(2)}=\kappa/2\pi$ must be obtainable from analysing the conical deficit in the Euclidean time direction at the outer horizon for \eqref{Kerr_Kruskal_4}. A detailed derivation of this is relegated to the Appendix-\ref{conical_deficit}, however an equivalent way to see this is to note that the conical deficit can also be perceived in the Lorentzian geometry by the $\{u,v\}$ light cone coordinates \eqref{LC_coord_Kerr_4} as the lack of affineness- which is indeed given by $\kappa$ in \eqref{kappa_Kerr_4} measured at $\rpl$.
\subsubsection{NHNext geometry, $\Ll=0$ to $\Ll \neq 0$}
\label{L_0_to_L_neq_0}
Let us compare the $NHNext$ metrics for the case obtained till now in literature $i.e.$ $\Ll=0$ to some allowed value of $\Ll\neq 0$. In either case the $NHNext$ metric takes the form \eqref{ads_4_ext} (\eqref{ads_4_ext} for the strict extremal case) in the near horizon coordinates $\{r,\tau,\theta,z\}$. However, the $\{\tau,z\}$ coordinates are relates differently to the Boyer-Lindquist coordinates at the conformal $AdS_\partial$ \eqref{t_phi_to_tau_z_Kerr_4}. Defining $\{\hat{\tau},\hat{z}\}$ as near horizon coordinates for $\Ll=0$ case and $\{\tau,z\}$ for $\Ll\neq 0$ we have
\be
\tau=(1-\mu\,\Ll)\hat{\tau}+\hat{z}\frac{\ell^2\Ll(1-\mu\,\Ll)}{a\,\Ll-\ell^2(1-\Omega_{\varphi}\Ll)},\hspace{0.4cm}z=\hat{z}\frac{\ell^2\Ll(1-\mu\,\Ll)}{a\,\Ll-\ell^2(1-\Omega_{\varphi}\Ll)}.
\label{near_hor_coor_change_Kerr_AdS_4}
\ee
It can be easily seen that the near horizon metric for $\Ll\neq 0$ cannot be obtained from the above (or any) coordinate change from a similar metric obtained with $\Ll=0$.  To see this first obtain the  near horizon metric for $\cL=0$  with $\{\hat{\tau},\hat{z}\}$ given by \eqref{t_phi_to_tau_z_Kerr_4} $i.e.$
\be
t=\hat{\tau},\hspace{0.3cm}\varphi=\hat{z}+\Omega_\varphi\hat{\tau}
\ee
and then use \eqref{strict_limits_ads_4} or \eqref{limits_ads_4}  with $\hat{\tau}\rightarrow \hat{T}/\lambda$ to obtain the near horizon metric. This would correspond to the near horizon metric used until now in literature. The near horizon coordinates $\{\hat{T},\hat{z}\}$ are then related to the $\cL\neq 0$ near horizon coordinates   $via$ \eqref{near_hor_coor_change_Kerr_AdS_4} with the $\tau$ \& $\hat{\tau}$ replaced with $T$ \& $\hat{T}$ respectively. However this change of coordinates would not result in the near horizon metric in \eqref{ads_4_ext} or \eqref{ads_4_next}. \emph{Note}, the near horizon metric \eqref{ads_4_ext} \& \eqref{ads_4_next} do not have explicit $\cL$ dependence, however the near horizon $z$ direction (and $T$ which is obtained from $\tau$) depends on the Boyer-Lindquist coordinate $via$ $\cL$.
Put simply, a simple coordinate change cannot take the near horizon metric \eqref{ads_4_next}(\eqref{ads_4_ext}) obtained for a particular value of $\Ll$ from the near extremal Kerr $AdS_4$ to a similar metric obtained for another value of $\Ll$. Therefore we seem to obtain a distinct IR geometry for each allowed value of $\Ll$.

\subsection{JT$_\Ll$ action $AdS_4$}
\label{Kerr_AdS_4_JT_der}
In this subsection we derive the JT action from the near extremal limit of metric \eqref{Kerr_Boyer_Lindquist_4} by dimensional reduction over an $S^2$ transverse to the  radial and temporal coordinates $\{r,\tau\}$. As noted above, we obtain a distinct IR geometry for each allowed value of $\Ll$. 
Since the JT model is derived using the such limits to the IR we label it as the JT$_\Ll$ model simply to distinguish the JT models obtained using different values of $\Ll$. As the near horizon metric is the same for any $\Ll$, this label simply indicates how the JT$_\Ll$ model is embedded in the full theory dual to the Kerr $AdS_4$. 
\\\\
We begin with the Euclideanized gravity action with a negative cosmological constant $-\Lambda_{4}=-\frac{3}{\ell^2}$
\be
I^{(E)}_{EH}=-\frac{1}{16\pi G_4}\int d^4x\sqrt{g_{(4)}}(R_{(4)}-2\Lambda_{(4)})-\frac{1}{8\pi G_4}\int d^3x\sqrt{h_{(4)}}K_{(4)}
\label{EH_ads_4}
\ee
and consider its dimensional reduction over the $S^2$  parametrized by $\{\theta,z\}$ for  the metric line elements of the form similar to the near horizon metrics \eqref{ads_4_ext}, \eqref{ads_4_next}
\be
ds^2_{(4)}=\cF(\theta)\frac{\Phi_0}{\Phi}ds^2_{(2)}+\Phi^2\cG(\theta) d\theta^2+\Phi^2\cG^{-1}(\theta)\sin^2\theta(dz+A)^2
\label{dim_red_anzats_Kerr_4}
\ee
The resulting action for the 2d metric $g_{(2)}$, the dilaton $\Phi$ and the $U(1)$ gauge field is 
\bea
I_{EH}^{(E)}&=&-\frac{1}{4G_4}\int\!\!\sqrt{g}\,\,d^2x\left(\Phi^2 R_{(2)}+\Lambda_{(4)}V_1\Phi+\frac{V_{-1}}{2\Phi}+\tfrac{1}{2}V_F\Phi^5 F_{\mu\nu}F^{\mu\nu}\right)+\cr&&\cr
&&-\frac{1}{2G_4}\int_\partial\!\!\sqrt{h_{(2)}}\,dx\,\left(\Phi^2 K_{(2)}-V_F\frac{\Phi^5}{\Phi_0}n_\mu F^{\mu\mu}A_\nu\right)\cr&&\cr
{\rm where}\hspace{0.2cm}V_1&=&\frac{8 \sqrt{2} \rpl^3 \ell ^5}{\sqrt{\ell ^2-3 \rpl^2} \left(18 \rpl^2 \ell ^2+3 \ell ^4-9 \rpl^4\right)}\hspace{0.3cm}V_F=\frac{\sqrt{2} \left(-3 \rpl^4+6 \rpl^2 \ell ^2+\ell ^4\right)}{\rpl^3 \ell ^3 \sqrt{\ell ^2-3 \rpl^2}},\cr&&\cr
\hspace{0.2cm}V_{-1}&=&\frac{4 \sqrt{2} \rpl^3 \ell ^3 \left(-3 \rpl^4-6 \rpl^2 \ell ^2+\ell ^4\right)}{\sqrt{\ell ^2-3 \rpl^2} \left(9 \rpl^6-21 \rpl^4 \ell ^2+3 \rpl^2 \ell ^4+\ell ^6\right)}.
\label{dim_red_EH_kerr_4}
\eea
Here we have added the extra boundary term to be in canonical ensemble $wrt$ the charge associated with the $U(1)$ symmetry. The $U(1)$ gauge symmetry corresponds to the $U(1)$ axi-symmetry direction of the Kerr black hole for the prevalent near horizon limits in literature. In principle this $U(1)$ symmetry is the symmetry of the line element \eqref{dim_red_anzats_Kerr_4} in the $z$ direction and this in turn depends on the value of the specific angular momenta $\Ll$ used in the change of coordinates $\{t,\varphi\}\rightarrow\{\tau,z\}$ in \eqref{t_phi_to_tau_z_Kerr_4}.
\\\\
The field strength $F_{\mu\nu}$ can be solved for exactly in 2d from its $eom$ 
\be
\nabla_\mu( \Phi^{5}F^{\mu\nu})=0\implies F_{\mu\nu}=\frac{i\sqrt{g_{_{(2)}}}Q}{\Phi^{5}}\epsilon_{\mu\nu}\implies F^2=-\frac{2Q^2}{\Phi^{10}}.
\label{Q_U_1_Kerr_4}
\ee
Substituting the above value for the $F^2$ we have
\bea
I_{EH}^{(E)}&=&-\frac{1}{4G_4}\int\!\!\sqrt{g_{_{(2)}}}\,\,d^2x\left(\Phi^2 R_{(2)}+\Lambda_{(4)}V_1\Phi+\frac{V_{-1}}{2\Phi}-V_F\frac{Q^2}{\Phi^5} \right)-\frac{1}{2G_4}\int_\partial\!\!\sqrt{h_{_{(2)}}}\,dx\,\Phi^2 K_{(2)}
\label{dim_red_EH_Q_Kerr_4}
\eea
The near horizon line elements \eqref{ads_4_ext} and \eqref{ads_4_next} are solutions to the above action. Therefore we can read off the value of the dilaton $\Phi=\Phi_0$ and $Q$ from the extremal\footnote{We can use the near extremal near horizon metric \eqref{ads_4_next} too provided we replace all instances of $\rpl\rightarrow r_0$.} near horizon metric \eqref{ads_4_ext}.
\\\\
The electric charge $Q$  and the dilaton $\Phi_0$ take the values
\be
Q^2=\frac{8 r_0^{10} \ell ^{10} \left(-2 r_0^2 \ell ^2+3 r_0^4-\ell ^4\right)}{\left(3 r_0^2-\ell ^2\right){}^3 \left(6 r_0^2 \ell ^2-3 r_0^4+\ell ^4\right){}^2},\hspace{0.5cm}\Phi_0^2=\frac{2r_0^2\ell^2}{\ell^2-3r_0^2}
\label{ext_Q_Kerr_4}
\ee
respectively.
We now expand the above action \eqref{dim_red_EH_Q_Kerr_4} for fluctuations of the dilaton about $\Phi_0$ as $\Phi=\Phi_0(1+\phi)$ upto linear orders in $\phi$ to obtain
\bea
I^{(E)}_{EH}&=&I_{(0)}+I_{JT}\cr&&\cr
I_{(0)}&=&-\frac{\Phi_0^2}{4G_4}\left(\int\!\!d^2x\sqrt{g_{_{(2)}}}\,R_{(2)}+2\int_\partial\!dx\,\sqrt{h_{_{(2)}}}K_{(2)}\right)\cr&&\cr
I_{JT}&=&-\frac{\Phi_0^2}{2G_4}\left[\int\!\!d^2x\,\sqrt{g_{_{(2)}}}\,\,\phi\left(R_{(2)}+2\right)+2\int_\partial\!dx\, \sqrt{h_{_{(2)}}}\,dx\,\phi K_{(2)} 	\right]
\label{JT_Kerr_4}
\eea
We clearly see that the 2d Newton's constant is entirely determined by $G_4$ and the extremal value of $\Phi_0$. Further the $eom$ for $\phi$ fixes $R_{(2)}=-2$ which is indeed the case for the $\{R,T\}$ components (within the box brackets) of the near horizon metric line elements  \eqref{ads_4_ext} \& \eqref{ads_4_next}. 

\subsection{Thermodynamics}
\label{Kerr_AdS_4_Thermodynamics}
We would first like to understand what canonical ensemble we would be analysing given the near horizon limits explored previously. As the JT  theory is obtained from dimensional reduction about the near horizon metric,  the charges associated with the symmetries in the transverse direction would be held fixed while analysing its thermodynamic response to any perturbation. In the near horizon metric \eqref{ads_4_ext} \& \eqref{ads_4_next} this is the $z$ direction and translations in this coordinate are generated by
\be
\partial_z=\frac{\partial_\varphi+\cL(\partial_t-\frac{a}{\ell^2}\partial_\varphi)}{1-\mu\,\cL}.
\ee
Given that $Q[-\partial_\varphi]\sim J$ and $Q[(\partial_t-\frac{a}{\ell^2}\partial_\varphi)]\sim M$ ($c.f.$  \cite{Papadimitriou:2005ii}) we have
\be
Q[\partial_z]\sim J-\cL \,M =const.
\ee
This should not be surprising as any perturbation following the geodesic \eqref{geodesic_cond_Kerr_4} with specific angular momentum $\cL$ would have the same relation between its charges.
The thermodynamic process we are therefore interested in is perturbing an extremal Kerr black hole with ADM Mass $M_0$ and angular momentum $J_0$ such that the resultant near extremal  black hole has an excess mass and angular momentum $\Delta M$ and $\Delta J$ respectively. Further we demand that 
\be
\Delta J=\Ll\,\Delta M
\label{Delta_J_Kerr_4}
\ee
This is indeed to be viewed as a perturbation due to an in-falling rotating null ray\footnote{We could take the particle to be a test particle released from the $AdS$ boundary in which case its trajectory at very late times $(\gg\beta)$ would be that of a null particle close to the horizon.} with specific angular momentum $\cL$. 
For values of $\Ll<\mu^{-1}_0$ we see that the resultant geometry would not be extremal and would thus have a small temperature which $T_H$. The above process can be imagined as caused by an in-falling perturbation with an energy $\Delta M$ measured at the $AdS_4$ boundary by stationary observer with a specific angular momentum $\Ll=\Delta J/\Delta M$. For such a process it was observed in \cite{Malvimat:2022oue} that the mutual information $I[A:B]$ between identical hemispheres $A$ \& $B$ in the $left$ and $right$ CFTs corresponding to the dual TFD state experience a disruption at a rate governed by the Lyapunov index
\be
\lambda_L=\frac{2\pi T_H}{(1-\mu\,\Ll)},\hspace{0.4cm}{\rm where}\,\,{\Ll<\mu^{-1}}
\label{Lyapunov_MI_Kerr_4}
\ee 
even at late times. Given the above constraint \eqref{Delta_J_Kerr_4} and the first law of black hole thermodynamics
\be
\delta M = T_H\, \delta\cS +\mu\,\delta J
\label{I_law_kerr_4}
\ee
we have the relation 
\be
\delta M=\frac{T_H}{(1-\mu\,\Ll)}\delta S
\label{I_law_plus_Delta_J_constraint_kerr_4}
\ee
where we associate $\delta$ with small fluctuations. Given the ADM mass $M$ and entropy $\cS$
\be
M=\frac{m}{G_4\Xi^2},\hspace{0.4cm}\cS=\frac{\pi(\rpl^2+a^2)}{G_4\Xi}
\label{M_S_Kerr_4}
\ee
and the above constraint on fluctuations \eqref{Delta_J_Kerr_4}, about an extremal configuration the excess mass and entropy are related to the small temperature as 
\be
\Delta M=\frac{4\pi^2 r_0^3\ell^4(\ell^2-r_0^2)T_H^2}{(\ell^2-3r_0^2)(\ell^4+6r_0^2\ell^2-3r_0^2)(1-\mu_0\,\Ll)},\hspace{0.3cm}\Delta S=\frac{8\pi^2 r_0^3\ell^4(\ell^2-r_0^2)T_H}{(\ell^2-3r_0^2)(\ell^4+6r_0^2\ell^2-3r_0^2)},\hspace{0.3cm}
\label{Delta_M_S_kerr_4}
\ee
were $\mu_0=\frac{\sqrt{(\ell^2-r_0^2)(\ell^2+3r_0^2)}}{2r_0\ell}$,
which confirms the first law \eqref{I_law_plus_Delta_J_constraint_kerr_4} for small variations of the above quantities with temperature. We intend to recover the above values of excess mass and entropy from a near horizon $JT$ analysis.

\subsubsection{NHNext JT$_\Ll$}
\label{Kerr_AdS_4_NHNext_JT_analysis}
As outlined in the introductory subsection-\ref{Intro_JT}, The non-dynamical topological piece $I_{(0)}$ in \eqref{JT_Kerr_4} can be used to determine the extremal value of the entropy $\cS$ by evaluating it on-shell.
To begin with this computation first write out a solution to the JT metric $eom$ as 
\be
ds^2_{(2)}=\frac{dR^2}{(R^2-\delta\rpl^2)}+(R^2-\delta\rpl^2)dT^2
\label{2d_near_ext_Euclid_metric}
\ee
which is the Euclidean version of the $\{R,T\}$ components in box bracket of the near extremal near horizon metric obtained in \eqref{ads_4_next}. 
Evaluating the $I_{(0)}$ for the above metric we have
\bea
I_{(0)}&=&-\frac{\Phi_0^2}{4G_4}\left(\int\!\!d^2x\sqrt{g}\,R_{(2)}+2\int_\partial\!dx\,\sqrt{h}K_{(2)}\right)\cr&&\cr
&=&-\frac{\Phi_0^2}{4G_4}\left[\int_{\delta\rpl}^{R_\infty}\hspace{-0.3cm}\int_0^{\beta^{(2)}}\hspace{-0.5cm}d^2x\sqrt{g}\,R_{(2)}+2\int_0^{\beta^{(2)}}\hspace{-0.5cm}dT\,\,\sqrt{h}\,K_{(2)}\right]\cr&&\cr
&=&\frac{\Phi_0^2}{4G_4}2\delta\rpl\beta^{(2)}=\frac{\pi\Phi_0^2}{G_4}=\cS_{ext}
\label{topological_S_ext_Kerr_4}
\eea
which is precisely $\frac{V^{(0)}_H}{4G_4}$ with $V^{(0)}_H=4\pi\Phi^2_0$ being the horizon area. Here we denoted $T_H^{(2)}=(\beta^{(2)})^{-1}=\frac{\delta\rpl}{2\pi}$ as the teperature of the 2d Euclidean black hole. Note that the above verification is independent of the value of $T_H^{(2)}$.
\\\\
We next turn to evaluating the excess entropy over extremality using $I_{JT}$. For this we would need the on-shell behaviour of the dilaton over the $AdS_2$ background \eqref{2d_near_ext_Euclid_metric} its value at the $AdS_2$ horizon.  
The static solution to the dilaton $\phi$ is 
\be
\phi= R\phi_\partial\implies\phi|_{\delta\rpl}=\delta\rpl\phi_\partial
\label{static_dilaton_ads_2_hor_Kerr_4}
\ee
The $AdS_2$ horizon at $R=\delta\rpl$ is also the location of the near extremal horizon as seen in the limit \eqref{strict_limits_ads_4}. To make contact with the higher dimensional near extremal geometry we need to specify 2 geometric data as mentioned in subsecttion-\ref{Intro_JT}. The first relates the $AdS_2$ temperature $T_H^{(2)}$ to $T_H$ \eqref{T_2_T_H_Kerr_4}. The second is how the fluctuation of the transverse volume at the horizon  relates to $\phi$. The latter is obtained by taking the limit ($\lambda\rightarrow 0$) \be
\rmi\rightarrow r_0-\lambda\cA\delta\rpl,\hspace{0.3cm}\rpl\rightarrow r_0+\lambda\cA\delta\rpl\,\,{\rm where}\,\,\mathcal{A}=\frac{2r_0^2\ell^2 \left(r_0^2+\ell ^2\right) \left(1-\mu_0\,\Ll\right)}{6 r_0^2 \ell ^2-3 r_0^4+\ell ^4}
\ee
used to obtain the near horizon metric \eqref{ads_4_ext} on  
\be
\frac{V_H}{4\pi}=\Phi^2,\hspace{0.4cm}{\rm where}\,\,V_H={\it Vol}(S^2)
\ee
of the full Kerr metric \eqref{Kerr_Boyer_Lindquist_4}.
$\phi_\partial$ can be then inferred from the transverse volume fluctuation $\Delta\Phi$ at the outer horizon  as
\bea
\Phi^2 &=& \Phi^2_0(1 +2\phi|_{\delta\rpl})=\Phi^2_0+2\Delta \Phi\cr&&\cr
\implies\hspace{0.3cm}\phi_\partial&=&\frac{\Delta \Phi}{\Phi_0\,\delta\rpl}=\frac{\Delta \Phi}{\Phi_0}\frac{(1-\mu_0\,\Ll)}{4\pi T_H}
\label{diltn_bndry_val_trns_vol_Kerr_4}
\eea
where we have used \eqref{static_dilaton_ads_2_hor_Kerr_4},\eqref{T_2_T_H_Kerr_4}. $\Delta\Phi$ is computed from the change in horizon volume linear in $\lambda$ as $\lambda\rightarrow 0$ of the black hole metric as given in \eqref{Kerr_Boyer_Lindquist_4} about extremality for $r_\pm=r_0\pm \lambda\,\cA\,\delta\rpl$ with $\cA$ as in \eqref{strict_limits_ads_4}
\be
\frac{\Delta\Phi}{\Phi_0}=\frac{r_0\ell^2(\ell^2-r_0^2)}{\ell^4+6r_0^2\ell^2-3r_0^2}\delta\rpl
\ee
The change in entropy is then simply computed  using the value of $\phi_\partial$ obtained above as the change on $V_H$ as
\be
\frac{\Delta \cS}{\cS_{ext}}=2\phi|_{R=\delta\rpl}=\delta\rpl\,\phi_\partial
\label{Delta_S_by_S_ads_2_Kerr_4}
\ee
which matches that computed form \eqref{Delta_M_S_kerr_4}.
\\\\
The change in mass above extremality is computed by simply computing the ADM mass in the 2d theory for the black ground \eqref{2d_near_ext_Euclid_metric} and the static dilaton \eqref{static_dilaton_ads_2_hor_Kerr_4}. In terms of the 2d parameters this takes the value
\be
M^{(2)}= \frac{\delta\rpl}{16\pi G_{(2)}}\phi|_{R=\delta\rpl}=\frac{\delta\rpl^2}{16\pi G_{(2)}}\phi_\partial,\hspace{0.5cm}{\rm where}\,\,\,G_{(2)}=\frac{G_{(4)}}{8\pi\Phi^2_0}
\label{M_ads_2_Kerr_5}
\ee
Substituting the value of $\phi_\partial$ and $\delta\rpl$ from \eqref{diltn_bndry_val_trns_vol_Kerr_4} and \eqref{T_2_T_H_Kerr_4} we find
\be
M^{(2)}=\frac{T_H^{(2)}}{2}\Delta\cS=\Delta M
\label{M_2_Delta_M-Kerr_4}
\ee
which matches with \eqref{Delta_M_S_kerr_4}. Thus we find that the excess mass and entropy are captured by the near horizon thermodynamics of the JT$_\Ll$ model including the first law for small fluctuations of $\delta M$ and $\delta\cS$ near extremality.
\be
\delta M=\frac{T_H}{(1-\mu\,\Ll)}\delta\cS
\label{first_law_JT_kerr_4}
\ee 
\section{Kerr in AdS$_\textbf{5}$ with $J_{\phi_1}=J_{\phi_2}$}
\label{MP_Kerr_AdS_5}
In this section we consider the analogous case for a Kerr black hole in $AdS_5$ but wit equal angular momentum. We consider the thermodynamics related to perturbing such an extremal black hole such that the resultant near extremal geometry has excess mass $\Delta M$ and excess angular momenta $\Delta J_{\varphi_1}$ \& $\Delta J_{\varphi_2}$ such that
\be
\Delta J_{\varphi_1}=\Llo\,\Delta M,\hspace{0.4cm}\Delta J_{\varphi_2}=\Llt\,\Delta M
\label{Delta_J_Kerr_MP_5}
\ee 
We proceed along similar lines as for the near extremal JT description of Kerr $AdS_4$ in the previous section to understand the thermodynamics from the near horizon JT perspective.
\\\\
We begin with briefly introducing the geometry of the Kerr black hole in $AdS_5$ with equal angular momentum, also known as Myers-Perry(MP) black holes.  Such a black hole  with the 2 angular momenta  equal $J_{\phi_1}=J_{\phi_2}=J$ preserves an $S^2$ fibration of $S^3$, therefore the geometry posses an $SO(3)\subset SU(2)\times SU(2)$ isometry. 
Hence, as compared to the generic Kerr black hole in $AdS_5$ the metric line element is much simpler.
In Boyer-Lindquist coordinates this takes the form
\bea
ds^2&=&\frac{r^2dr^2}{(r^2+a^2)\Delta(r)}\,-\,\frac{1}{\Xi}\Delta(r)e^{U_2-U_1}dt^2\,+\,e^{-U_1}d\Omega_2^2\,+\,e^{-U_2}\left(\sigma^3+A\right)^2,\cr&&\cr&&\cr
\hspace{-2cm}{\rm where}\hspace{3cm}&&\Delta(r)=1+\frac{r^2}{l^2}-\frac{2mr^2}{(r^2+a^2)^2},\hspace{0.3cm}
\Xi=1-\frac{a^2}{l^2},\cr&&\cr
&&e^{-U_2}=\frac{r^2+a^2}{4\,\Xi}+\frac{2ma^2}{2\,\Xi^2(r^2+a^2)},\hspace{0.3cm}e^{-U_1}=\frac{r^2+a^2}{4\,\Xi},\cr&&\cr
&&A=\frac{a}{2\,\Xi}\left(\frac{r^2+a^2}{l^2}-\frac{m}{r^2+a^2}\right)e^{U_2}dt	
\label{Kerr_Boyer_lindquist_5_MP}
\eea
The angular line elements are
\be
\sigma^3=d\psi+\cos \theta\,d\varphi,\hspace{0.4cm}d\Omega^2_2=d\theta^2+\sin^2\theta\,d\varphi^2
\ee
where the $S^2$ symmetry is manifest. The polar coordinates for $S^3$ range from $\theta\in \{0,\pi\}$ ,  $\varphi\in\{0,2\pi\}$ and $\psi\in\{0,4\pi\}$\footnote{The toric coordinates describing the $S^3$ are $\eta=\theta/2$ and $\varphi_1=(\psi-\varphi)/2$, $\varphi_2=(\psi+\varphi)/2$ where \\$d\Omega_3^2=d\eta^2+\sin^2\theta \,d\varphi_1+\cos^2\theta \,d\varphi_2$.   }.
The axi-symmetry directions of the black hole are $\varphi_1=(\psi-\varphi)/2$ and $\varphi_2=(\psi+\varphi)/2$. The black hole is defined by the mass parameter $m$ and the angular momentum parameter $a$ with
the ADM mass and 2 equal angular momenta given by \cite{Skenderis:2006uy}
\be
M=M_C+\frac{2\pi^2 m(3+\frac{a^2}{l^2})}{\Xi^3},\hspace{0.3cm}
J=\frac{8\pi^2 m a}{\Xi^3},\hspace{0.3cm}
M_C=\frac{3\pi^2l^2}{4}
\label{M_M_cas_J_Kerr_MP_5}
\ee
Here  the Casimir energy $M_C$ associated with $S^3$ is independent of the black hole parameters.
The entropy and temperature is
\be
\mathcal{S}=\frac{\pi^2(\rpl^2+a^2)^2}{2G_N\rpl \Xi^2},\hspace{0.5cm}T_H=	\frac{\rpl^2-a^2+\frac{2\rpl^4}{l^2}}{2\pi\rpl(\rpl^2+a^2)}
\label{T_H_S_Kerr_MP_5}
\ee
with $l$ and $G_N$ denoting  the $AdS_5$ radius  and the 5d Newton's constant.
The Boyer-Lindquist coordinates used above are adapted from that of flat space, in asymptotically $AdS_5$ space these have a non-zero angular velocity at the boundary about $\phi_1$ and $\phi_2$. The  horizon angular velocity with respect to an stationary boundary frame is given by
\be
\Omega=\frac{a\left(1+\frac{\rpl^2}{l^2}\right)}{\rpl^2+a^2}=\mu
\label{mu_Kerr_MP_5}
\ee
and this plays the role of chemical potential $\mu$ relevant for the thermodynamics of the black hole. The reader is referred to \cite{Skenderis:2006uy} for a detailed derivation of asymptotic charges relevant for thermodynamics for generic Kerr black holes in $AdS_5$.


\subsection{Near Horizon limits}
\label{MP_Kerr_AdS_5_NH_limits}
We proceed first by generalizing the near horizon limits found in literature by rewriting the metric along in-out going null geodesies. As seen in \cite{Malvimat:2022fhd} it is along such geodesies that the black hole is perturbed. We write the null geodesic vector field $\xi^\mu\partial_\mu$ as 
\be
\xi^2=0,\hspace{0.3cm}g_{\mu\nu}\xi^\mu \zeta^\nu_t=\mathcal{E}=1,\hspace{0.3cm}
g_{\mu\nu}\xi^\mu\zeta^\nu_{\varphi_1}=\mathcal{L}_{\varphi_1},\hspace{0.3cm}g_{\mu\nu}\xi^\mu\zeta^\nu_{\varphi_2}=\mathcal{L}_{\varphi_2}
\label{geodesic_eq_Kerr_5_MP}
\ee
where the charges are defined along the killing isometries 
\bea
&&\zeta_{\varphi_1}=\partial_{\varphi_1}, \hspace{0.3cm}\zeta_{\varphi_2}=\partial_{\varphi_2}, \hspace{0.3cm}\zeta_t=\partial_t-\frac{2a}{l^2}\partial_\psi\cr&&\cr
&&{\rm where}\hspace{0.3cm}\varphi_1=\frac{\psi-\varphi}{2},\hspace{0.3cm}\varphi_2=\frac{\psi+\varphi}{2}
\label{KV_Kerr_MP_5}
\eea
Note in the above coordinates only the $\psi$ direction has a boundary velocity.
Kerr black holes have an additional higher-form symmetry in terms of Killing-Yano tensor $f_{\mu\nu}=-f_{\nu\mu}$, the conserved quantity corresponding to this is the Carter's constant and is defined using symmetric Killing tensor  $K_{\mu\nu}$ which satisfies
\be
\nabla_\rho K_{\mu\nu}=0,\,\,\,K_{\mu\nu}=K_{\nu\mu},\,\,\,K_{\mu\nu}=-f_\mu^{\,\,\,\sigma}f_{\sigma\nu}
\ee
The Carter's constant is then defined as
\be
\xi^\mu\xi^\nu K_{\mu\nu}=\mathcal{Q}
\ee  
The Killing tensor for generic Kerr $AdS_5$ is worked out in \cite{Kunduri:2005fq}. We make use of it in our coordinates for the case of equal angular momentum.  We define in-out going null pairs denoted by $\xi^{\mu}_\pm\partial_\mu$ where the $\xi^\mu_-$ is obtained from $\xi^\mu_+$ by reversing the signs of energy $\mathcal{E}$, $\Ll_{\varphi_1}$ \& $\Ll_{\varphi_2}$ and $\xi^\theta_+$  and then putting $\mathcal{E}=1$. 
\\\\
We would like to associate light-cone coordinates along the one-forms dual to $\xi_\pm\cdot\partial$
\be
\xi^+_\mu dx^\mu =du,\hspace{0.3cm}\xi^+_\mu dx^\mu =dv
\label{dLC_Kerr_MP_5}
\ee
This implies exactness which in turn implies 
\be
\partial_\theta\xi^\pm_r=\partial_r\xi^\pm_\theta=0,\hspace{0.4cm}\implies\mathcal{Q}=const 
\ee
We choose the value of $\cQ$ such that the geodesies that starts out at the equator stays in the equatorial plane $i.e.$ $\dot{\theta}=\xi_\pm^\theta=0$ at $\theta=\pi/2$, this  implies holding $\mathcal{Q}$ constant at
\be
\mathcal{Q}=\Xi\,(2(\Llo^2+\Llt^2)-l^2).
\label{Q_equator_Kerr_MP_5}
\ee
Using the above one-forms we rewrite the line element \eqref{Kerr_Boyer_lindquist_5_MP} as
\be
ds^2=F(\xi^+_\mu dx^\mu)(\xi^-_\nu dx^\nu)+h_\varphi(d\varphi+h_\tau d\tau)^2 +h_z(dz+g_\varphi d\varphi+A_\tau d\tau)^2 + h_\theta(d\theta+g_\tau d\tau)^2
\ee
where we have defined the new coordinates $\{\tau, z\}$ as
\be
\tau=\left(1-\frac{a}{l^2}(\Llo+\Llt)\right) t + \frac{\Llo-\Llt}{2}\varphi-\frac{\Llo+\Llt}{2}\psi
\label{tau_to_t_phi_psi_Kerr_MP_5}
\ee
\bea
\psi&=&\eta\, z-\frac{2a(a^2-l^2)}{l^2(a^2+\rpl^2)-a(\rpl^2+l^2)(\Llo+\Llt)}\tau,\hspace{0.3cm}
\label{psi_to_z_tau_Kerr_MP_5}
\eta=\frac{1-\tfrac{a}{l^2}(\Llo+\Llt)}{1-\mu (\Llo+\Llt)}, \hspace{0.3cm}{\rm where\,\,\,\,} \mu=\Omega
\eea
Inverting the above relations we have
\bea
\psi&=&\frac{z \left(1-\frac{a \mathcal{L}_+}{\ell ^2}\right)+\frac{2 a   \left(\ell ^2-a^2\right)}{\ell ^2 \left(a^2+r_+^2\right)}\tau}{1-\mu\,  \mathcal{L}_+},\hspace{0.3cm}
t=\frac{ \tau +\tfrac{1}{2}z \,\mathcal{L}_+}{1- \mu \, \mathcal{L}_+}-\frac{\varphi\,\mathcal{L}_-  }{2 \left(1-\frac{a \mathcal{L}_+}{\ell ^2}\right)}
\label{t_psi_to_tau_z_Kerr_MP_5}
\eea
where $\Ll_\pm=\Llo\pm\Llt$.
The light cone coordinates can be used to define tortoise coordinate $r_*$ in the radial direction
\bea
&&du=\xi^+_\mu dx^\mu \implies u=r_*-\tau-\tilde{g}(\theta)\cr&&\cr
&&dv=\xi^-_\mu dx^\mu \implies v=r_*+\tau+\tilde{g}(\theta)
\label{LC_coordinates_Kerr_MP_5}
\eea
where $r_*$ integrating $dr_*=\xi_rdr$ and choosing the boundary at $r_*=0$ and 
\be
\tilde{g}'(\theta)=\frac{\sqrt{\Llt^2-\Llo^2-(\Llo^2+\Llt^2)\cos\theta}}{\sin\theta}
\label{tilde_g_Kerr_MP_5}
\ee
which can be integrated $w.r.t$ $\theta$.
We can see that these $\{u,v\}$ coordinates are not smooth at the horizons; for instance for the vector field $\chi$ that generates translations along $v$ $i.e.$ 
$\chi= \partial_u$ we find $\chi\cdot\nabla\chi=\mathcal{K}\chi$ where
 \be
\mathcal{K}=\frac{1}{2}\xi^+\cdot\partial F, \hspace{0.3cm}{\rm with}\,\,\mathcal{K}|_{r=\rpl}=\kappa\neq 0
 \ee
 \be
\kappa=\frac{\kappa_0}{(1-\mu\,\Ll_+)},\hspace{0.7cm}{\rm where}\,\,\,\, \kappa_0=2\pi T_H,\,\,\mu=\Omega
\label{kappa_Kerr_MP_5}
\ee
Here the non-vanishing of $\mathcal{K}$ at the outer horizon signals the non-smoothness of the $\{u,v\}$ coordinates. 
One can therefore define the smooth Kruskal coordinates $\{U,V\}$ as before
\be
U=-e^{\kappa(r_*-\tau-\tilde{g}(\theta))},\,\,V=e^{\kappa(r_*+\tau+\tilde{g}(\theta))},\hspace{0.7cm}{\rm where} \,\,\kappa=\mathcal{K}\vert_{\rpl}
\label{Kruskal_coordinates_equator_Kerr_MP_5}
\ee
Using the above Kruskal coordinates we can rewrite the metric line element \eqref{Kerr_Boyer_lindquist_5_MP} as
\bea
&&ds^2=\frac{F}{\kappa^2 UV}dUdV + h_z(dz+g_\varphi d\varphi+A_\tau d\tau)^2 +h_\varphi(d\varphi+h_\tau d\tau)^2  + h_\theta(d\theta+g_\tau d\tau)^2\cr&&\cr
{\rm with}&&d\tau=\frac{1}{2UV}(UdV-VdU)-\tilde{g}'(\theta)d\theta
\label{metric_kruskal_L_Kerr_MP_5}
\eea
The above coordinates $\{U,V,\theta,\varphi,z\}$ or $\{r,\tau,\theta,\varphi,z\}$ are well suited for the near horizon region as seen by null geodesies with specific angular momenta $\Llo$ \& $\Llt$ along the axi-symmetry directions of the black hole. It is only along these coordinates that the outer horizon appears smooth and devoid of any coordinate singularity. If one were to Euclideanize the above line element by taking $\tau\rightarrow i\tau_E$ then  demanding that the Euclidean geometry be smooth or devoid of conical deficit at the outer horizon would imply a periodicity of  $(\kappa/(2\pi))^{-1}$ for $\tau_E$. This is consistent 
with the Kruskal exponent being $\kappa$ in \eqref{Kruskal_coordinates_equator_Kerr_MP_5} in the Lorentzian picture as both these conditions stem from the geometric requirement of the metric being smooth at the outer horizon. Therefore when one takes the near horizon limit in the above coordinates it is as though the outer horizon is being approached along the family of rotating null geodesies along $\{U,V\}$.
\\\\\ 
The strict extremal limit can be obtained by taking the $\lambda\rightarrow 0$ limit about the extremal horizon $r_0$ as
\bea
&&\rmi=r_0-\lambda\, \mathcal{A}\,\delta\rpl,\hspace{0.3cm}\rpl=r_0+\lambda\, \mathcal{A}\,\delta\rpl,\hspace{0.3cm} r=r_0+\lambda\,\mathcal{A}R,\hspace{0.3cm}\tau=\frac{T}{\lambda}\cr&&\cr
&&{\rm where}\hspace{0.3cm}\mathcal{A}=\frac{r_0^2(r_0^2+\ell^2)(1-\mu\,\Ll_+)}{(4r_0^2+\ell^2)}
\label{strict_ext_5_MP}
\eea
The near horizon metric then takes the same form  
\bea
ds^2_{AdS_5(\rm NHext)}&=&\mathcal{F}_{r_0}\left[\frac{dR^2}{R^2}-R^2dT^2\right]+\mathcal{G}_{r_0}\,(d\theta^2+\sin^2\theta\, d\varphi^2)
+\mathcal{H}_{r_0}\left(dz+g_\varphi d\varphi+A_T\,dT\right)^2\cr&&\cr
{\rm where}&&\mathcal{F}_{r_0}=\frac{r_0^2\ell^2}{2(4r_0^2+\ell^2)},\hspace{0.3cm}\mathcal{G}_{r_0}=\frac{r_0^2\ell^2}{2\ell^2-4r_0^2},\hspace{0.3cm}\mathcal{H}_{r_0}=\frac{r_0^2\ell^4}{(\ell^2-2r_0^2)^2}\cr&&\cr
&&A_{r_0\,T}=-\frac{\left(\ell ^2-2 r_0^2\right) \sqrt{2 r_0^2+\ell ^2}}{4 r_0^2 \ell +\ell ^3}R 
\hspace{0.3cm}g_\varphi=\cos\theta + \frac{\Ll_- \left(2 r_0^2-\ell ^2\right) \sqrt{2 r_0^2+\ell ^2}}{2 r_0 \left(r_0 \Ll_+ \sqrt{2 r_0^2+\ell ^2}-\ell ^3\right)}
\label{ext_ads_Kerr_MP_5Kerr}
\eea
The above metric is the same as in \cite{Lu:2008jk} for $\Ll_-=0$.
The simultaneous near extremal near horizon limit of the Kerr $AdS_5$ geometry can be reached by scaling the black hole horizons and the $\{r,\tau\}$ coordinates as
\bea
&&\rmi=\rpl-2\lambda\, \mathcal{A}\,\delta\rpl,\hspace{0.3cm} r=\rpl+\lambda\,\mathcal{A}(R-\delta\rpl),\hspace{0.3cm}\tau=\frac{T}{\lambda}\cr&&\cr
&&{\rm where}\hspace{0.3cm}\mathcal{A}=\frac{\rpl^2(\rpl^2+\ell^2)(1-\mu\,\Ll_+)}{(4\rpl^2+\ell^2)}
\label{limits_ads_MP_5}
\eea
Here the relative scaling factor $\mathcal{A}$ is needed so that the near horizon metric takes the form in which the space-time in the $\{R,T\}$ coordinates is an $AdS_2$ Schwarzchild 
\bea
ds^2_{AdS_5(\rm next)}&=&\mathcal{F}\left[\frac{dR^2}{(R^2-\delta\rpl^2)}-(R^2-\delta\rpl^2)dT^2\right]+\mathcal{G}\,(d\theta^2+\sin^2\theta\, d\varphi^2)\cr&&\cr
&&\hspace{0.6cm}+\mathcal{H}\left(dz+g_\varphi d\varphi+A_T\,dT\right)^2\cr&&\cr
{\rm where}&&\mathcal{F}=\frac{\rpl^2\ell^2}{2(4\rpl^2+\ell^2)},\hspace{0.3cm}\mathcal{G}=\frac{\rpl^2\ell^2}{2\ell^2-4\rpl^2},\hspace{0.3cm}\mathcal{H}=\frac{\rpl^2\ell^4}{(\ell^2-2\rpl^2)^2}\cr&&\cr
&&A_T=-\frac{\left(\ell ^2-2 \rpl^2\right) \sqrt{2 \rpl^2+\ell ^2}}{4 \rpl^2 \ell +\ell ^3}(R-\delta\rpl) 
\hspace{0.3cm}g_\varphi=\cos\theta + \frac{\Ll_- \left(2 \rpl^2-\ell ^2\right) \sqrt{2 \rpl^2+\ell ^2}}{2 \rpl \left(\rpl \Ll_+ \sqrt{2 \rpl^2+\ell ^2}-\ell ^3\right)}\cr&&
\label{Next_AdS_Kerr_MP_5}
\eea
One can replace $\rpl\rightarrow r_0$ in the above coefficients to the leading order in $\lambda$ as $\rpl=r_0+\lambda\cA\delta\rpl$ which is the same as  the near horizon metric obtained in \cite{Lu:2008jk} for $\Ll_-=0$ .
\\\\
It is worth noting that unlike the near horizon coordinates used in literature the above near horizon time $T$ is not scaled with any constant factor relative to $\tau$  which for $\cL_\pm=0$ is the Boyer-Lindquist time. We can rewrite the above metric in the form \eqref{dim_red_gen_Kerr_MP_5} to be used for obtaining the JT action \eqref{JT_ads_action_Kerr_MP_5}
\bea
&&ds^2_{AdS_5(\rm NHNext)}=\frac{\cF_0}{\Sigma_0\Phi_0^{3/2}}\left[\frac{dR^2}{(R^2-\delta\rpl^2)}-(R^2-\delta\rpl^2)dT^2\right]+\frac{\Phi_0^3}{\gamma_0\Sigma_0}d\Omega^2_{(2)}+\Sigma^2_0(dz+g_\varphi d\varphi+A_TdT)^2\cr&&\cr
&&{\rm where}\hspace{2cm}\Sigma_0=\sqrt{\cH_{r_0}},\hspace{0.2cm}\Phi_0=(\gamma_0 \cG_{r_0})^{1/3}\cH_{r_0}^{1/6},\hspace{0.3cm}\cF_0=\frac{2\sqrt{\gamma_0}\cG_{r_0}^{5/2}\cH_{r_0}^{3/4}}{(3\cH_{r_0}-4\cG_{r_0})}
\label{Next_AdS_5_1}
\eea
The charge $Q$ associated with the gauge field $wrt$ the above 2d metric in $\{R,T\}$ is
\be
Q^2=\frac{\gamma_0^3\rpl^{15}\ell^{22}\,(2\rpl^2+\ell^2)}{8(\ell^2-2\rpl^2)^{10}(4\rpl^2+\ell^2)^2}
\label{Q_to_J_ext}
\ee
The 2d horizon at $R=\delta\rpl$ is also the location of the near extremal horizon at $\rpl=r_0+\lambda\cA \delta\rpl$, therefore we would need to relate it to the $T_H$ as seen in the higher dimensional near extremal geometry. Using similar arguments to those outlined in the previous section for the Kerr $AdS_4$ near horizon limits in \eqref{NH_ads_2_Kruskal_BL_Kruskal_Kerr_4},\eqref{NH_ads_2_Kruskal_BL_Kruskal_Kerr_4} and \eqref{T_2_T_H_Kerr_4} we identify 
\be
T^{(2)}_H=\frac{T_H}{(1-\mu\,\Ll_+)}
\label{T_H_T_H_Kerr_MP_5}
\ee
As in the previous case this identification is consistent with demanding smoothness in either the Lorentzian line element \eqref{metric_kruskal_L_Kerr_MP_5} along the $\{U,V\}$ directions, or demanding lack of a conical deficit in the Euclideanised version of \eqref{2d_near_ext_Euclid_metric} by $\tau\rightarrow i\tau_{E}$.
\subsubsection{NHNext geometry, $\Ll_\pm=0$ to $\Ll_\pm\neq 0$}
\label{L_pm_0_to_L_pm_neq_0}
Let us compare the $NHNext$ geometries obtained in literature $i.e.$ using $\Ll_\pm=0$, to that obtained for allowed values of $\Ll\neq 0$. In contrast with the previous case of Kerr $AdS_4$ in subsection-\ref{L_0_to_L_neq_0}, here even in terms of the near horizon coordinates $\{r,\tau,\theta,\varphi,z\}$ we find a difference in the near horizon metric \eqref{Next_AdS_Kerr_MP_5}(\eqref{ext_ads_Kerr_MP_5Kerr}) as compared to the $\Ll_\pm=0$ case due to the dependence on $\Ll_-$. This difference vanishes for the case of $\Ll_-=0$. This- one might believe, is not surprising as a perturbation with $\Ll_-\neq 0$ makes the black hole have un-equal angular momentum. 
\\\
However, it is worth pointing out that the extra symmetry in the $S^2$ parametrized by $\{\theta,\phi\}$ persists even when $\Ll_-\neq 0$ in \eqref{ext_ads_Kerr_MP_5Kerr} and \eqref{Next_AdS_Kerr_MP_5}. Independent of this feature of an explicit dependence on $\Ll_-$ in the near horizon geometry we can make a similar observation as in subsection -\ref{L_0_to_L_neq_0} for the Kerr $AdS_4$. The IR geometries obtained for different values of $\Ll_\pm$ from  MP Kerr $AdS_5$ \eqref{MP_Kerr_AdS_5} are not related in the IR by some simple coordinate change. We comment more on this in the Discussions \& Conclusions section. 

\subsection{JT$_\Ll$ action $AdS_5$}
\label{MP_Kerr_AdS_5_JT_der}
Next we derive the JT model for each of the above near horizon limits. As we obtain a distinct IR geometry for different values of $\Ll_\pm$ we label the JT model as JT$_\Ll$ - to indicate that this model embeds itself differently in the IR theory in the dual of MP Kerr $AdS_5$ for different values of $\Ll_\pm$.
\\\\
We begin with the Euclideanised Einstein-Hilbert action in 5d with negative cosmological constant $-\Lambda_{(5)}=-\frac{6}{\ell^2}$.
\be
I_{EH}^{(E)}=-\frac{1}{16\pi G_5}\int d^5x\sqrt{g_{(5)}}\left(R_{(5)}-2\Lambda_{(5)}\right)-\frac{1}{8\pi G_5}\int d^4x\sqrt{\gamma}K_{(5)}
\label{EH_ads_5}
\ee
Given the form of the near horizon metrics for the extremal \eqref{ext_ads_Kerr_MP_5Kerr} and the near extremal \eqref{Next_AdS_Kerr_MP_5} Kerr geometry we write the  ansatz for the near horizon metric as 
\be
ds^2_{(5)}=\frac{\mathcal{F}_0}{\Phi^{3/2}\Sigma}ds^2_{(2)}+ \frac{\Phi^3}{\gamma_0\Sigma}d\Omega^2_{(2)}+\Sigma^2(\sigma^3 +A)^2
\label{dim_red_gen_Kerr_MP_5}
\ee
where we scale the $S^2$ with a constant $\gamma_0$ which can be set to 1. The volume of the compact directions transverse to the 2d metric $ds^2_{(2)}$ is
\be
V_H=\frac{16\pi^2}{\gamma_0}\Phi^3
\label{transverse_volume_Kerr_MP_5}
\ee
which is the volume of the horizon at extremality. The action \eqref{EH_ads_5} for the above metric ansatz reduces to an action for the 2d metric $g_{(2)}$, dilaton $\Phi$ and the $U(1)$ gauge field.
\bea
I_{2d}^{(E)}&=&-\frac{\pi}{\gamma_0 G_5}\int d^2x\sqrt{g}\left(\Phi^3R-2\cF_0\frac{\Phi^{3/2}}{\Sigma}\Lambda_{(5)} +\cF_0 \frac{4\Phi^3-\gamma_0\Sigma^3}{2\Phi^{9/2}}+\frac{1}{4\cF_0}\Sigma^3\Phi^{9/2}F_{\mu\nu}F^{\mu\nu}
-\frac{3}{2}\frac{\Phi^3}{\Sigma^2}(\nabla\Sigma)^2
\right)+\cr&&\cr
&&-2\frac{\pi}{\gamma_0G_5}\int_\partial dT\left(\Phi^3 K-\frac{\Phi^{9/2}\Sigma^3}{2}n_\mu F^{\mu\nu}A_\nu\right)
\label{AO_AdS_5}
\eea
where we have added the necessary boundary term to be in the canonical ensemble.
Here the $U(1)$ gauge direction corresponds to the translational (rotational) symmetry along the $z$ coordinate. The directions transverse to $\{R,T\}$ constitute a 3 dimensional compact manifold which is an $S^2$ fibred over the $U(1)$ direction with the $\Sigma$ field determining this fibration. The $\Sigma$ field can be thought of as	determining the departures of the transverse direction from being an $S^3$ and hence has been termed as a squashing parameter \cite{Castro:2018ffi,Castro:2021csm}. 
Solving the gauge field $eom$ we find
\be
\nabla_\mu( \Sigma^3\,\Phi^{9/2}F^{\mu\nu})=0\implies F_{\mu\nu}=\frac{i\sqrt{g}\,\,Q}{\Sigma^3\,\Phi^{9/2}}\epsilon_{\mu\nu}\implies F^2=-\frac{2Q^2}{\Sigma^6\,\Phi^9}
\label{F_sol}
\ee
where $\epsilon_{\mu\nu}$ is the completely anti-symmetric matrix in 2d and $Q$ is the (real) charge associated with the gauge field. This charge which appears in determining the $U(1)$ field strength is related to the global charge $J$ associated with the axi-symmetric coordinate $\psi=\varphi_1+\varphi_2$ of the 5d black hole \eqref{Kerr_Boyer_lindquist_5_MP}. Substituting the above value for $F^2$ we have
\bea
I_{2d}^{(E)}&=&-\frac{\pi}{\gamma_0 G_5}\int d^2x\sqrt{g}\left(\Phi^3R-2\cF_0\frac{\Phi^{3/2}}{\Sigma}\Lambda_{(5)} +\cF_0 \frac{4\Phi^3-\gamma_0\Sigma^3}{2\Phi^{9/2}}-\frac{1}{2\cF_0}\frac{Q^2}{\Sigma^3\Phi^{9/2}}
-\frac{3}{2}\frac{\Phi^3}{\Sigma^2}(\nabla\Sigma)^2
\right)+\cr&&\cr
&&-2\frac{\pi}{\gamma_0G_5}\int_\partial dT\left(\Phi^3 K-\frac{\Phi^{9/2}\Sigma^3}{2}n_\mu F^{\mu\nu}A_\nu\right)
\label{AO_Fsq_sub_AdS_5}
\eea
Expanding the above action to linear order in the fluctuation of the fields $\Phi=\Phi_0(1+\phi)$ and $\Sigma=\Sigma_0(1+\sigma)$ we find that only linear fluctuations in $\phi$ are relevant as linear terms proportional to the fluctuations of the squashing field $\sigma$ vanish. 
\bea
&&I_{2d}^{(E)}=I_{(0)}+I_{JT},\cr&&\cr
{\rm where}\hspace{0.3cm}
&&I_{(0)}=-\frac{\pi}{G_5}\frac{\Phi_0^3}{\gamma_0}\left\lbrace\int d^2x\sqrt{g}\, \,R_{(2)}+2\int_\partial dx\sqrt{h}\,\,K_{(2)}\right\rbrace\cr&&\cr
&&I_{JT}=-\frac{\pi}{G_5}\frac{3\Phi_0^3}{\gamma_0}\left\lbrace\int d^2x\sqrt{g}\,\phi\left[ R_{(2)}+\Lambda_{(2)}\right]
+2\int_\partial dx\sqrt{h}\,\phi\,K_{(2)}\right\rbrace
\label{JT_ads_action_Kerr_MP_5}
\eea
Here $\Phi_0$ and $\Sigma_0$ take the extremal value in \eqref{Next_AdS_5_1}. For expansion about the extremal solution we find 
\bea
{\rm here}&&\hspace{0.2cm}\gamma_0^2=\frac{Q^2}{\cF_0^2\Sigma_0^6}-\frac{4\Lambda_{(5)}\Phi_0^6}{3\Sigma_0^4},\hspace{0.3cm}
\Lambda_{(2)}=2
\label{JT_ads_charge_Kerr_MP_5}
\eea
We need to add a holographic counter term proportional to the boundary value of the dilaton $\phi$ to make $I_{JT}$ on-shell finite.

\subsection{Thermodynamics}
\label{MP_Kerr_AdS_5_Thermodynamics}
As in the Kerr $AdS_4$ case the canonical ensemble described by the thermodynamic fluctuations of the JT theory are determined by  the charges associated with the symmetries in the transverse direction being held fixed.  In the near horizon metric \eqref{ext_ads_Kerr_MP_5Kerr} \& \eqref{Next_AdS_Kerr_MP_5} this is the $z$ direction and translations in this coordinate are generated by
\be
\partial_z=\frac{2\partial_\psi+\cL_+(\partial_t-2\frac{a}{\ell^2}\partial_\psi)}{2(1-\mu\,\cL)}.
\ee
Given that $Q[-2\partial_\psi]\sim J_{\varphi_1}+J_{\varphi_2}=J_+$ and $Q[(\partial_t-2\frac{a}{\ell^2}\partial_\psi)]\sim M$ ($c.f.$  \cite{Papadimitriou:2005ii} \footnote{\cite{Papadimitriou:2005ii} uses $\varphi_{1,2}$ coordinates as in \eqref{KV_Kerr_MP_5}.}) we have
\be
Q[\partial_z]\sim J_+-\cL_+ \,M =const.
\ee
This is consistent with any perturbation following the geodesic \eqref{geodesic_eq_Kerr_5_MP} with specific angular momentum $\cL_{\varphi_{1,2}}$ would have the same relation between its charges.
We are interested in the thermodynamics of perturbing the extremal Kerr black hole such that the resultant near extremal geometry has a different  angular momenta than the extremal one. This is identical to the process studied for generic MP Kerr black holes in $AdS_5$ with regards to the change in its Mutual Information between two identical hemispheres in the $left$ and $right$ CFTs corresponding to the dual TFD state \cite{Malvimat:2022fhd}.
We intend to perturb an extremal MP Kerr black hole such that its ADM mass changes by $\Delta M$ and its angular momenta change as
\be
\Delta J_{\varphi_1}=\Llo\Delta M \hspace{0.2cm}\& \hspace{0.2cm}\Delta J_{\varphi_2}=\Llt\Delta M,
\label{Delta_J_Kerr_MP_5}
\ee 
As we would perturb the 2 angular momenta differently we need to know how the ADM mass and angular momenta depend on the parameters ${a,b,m}$ for a generic Kerr black hole in $AdS_5$. These are
\bea
&&M=\frac{\pi}{8G_{(5)}}\frac{m(2\Xi_a+2\Xi_b-\Xi_a\Xi_b)}{\Xi_a^2\Xi_a^2},\hspace{0.5cm}M_{\rm Casimir}=\frac{3\pi\ell^2}{64 G_{(5)}}\left(1+\frac{(\Xi_a-\Xi_b)^2}{9\Xi_a\Xi_b}\right)\cr&&\cr
&&J_{\varphi_1}=\frac{\pi}{4G_{(5)}}\frac{m a}{\Xi_a^2\Xi_b},\hspace{0.5cm}J_{\varphi_2}=\frac{\pi}{4G_{(5)}}\frac{m b}{\Xi_a\Xi_b^2},\hspace{0.5cm}\mathcal{S}=\frac{\pi^2(\rpl^2+a^2)(\rpl^2+b^2)}{2G_{(5)}\rpl\Xi_a\Xi_b}\cr&&\cr
&&T_H=\frac{1}{2\pi	}\left(\rpl\left(1+\frac{\rpl^2}{\ell^2}\right)\left(\frac{1}{\rpl^2+a^2}+\frac{1}{\rpl^2+b^2}\right)-\frac{1}{\rpl}\right)\cr&&\cr
&&\mu_{\varphi_1}=\frac{a(\rpl^2+\ell^2)}{(a^2+\ell^2)\ell^2},\hspace{0.5cm}\mu_{\varphi_2}=\frac{b(\rpl^2+\ell^2)}{(b^2+\ell^2)\ell^2}.
\label{genric_kerr_ADM}
\eea
Given the constraint \eqref{Delta_J_Kerr_MP_5} we can determine how the parameters $a$ and $b$ change with regards to the change in the outer horizon radius $\tilde{\delta}\rpl$ when $a=b$
\bea
&&\delta a=\frac{\tilde{\delta}\rpl \left(a^2-\ell ^2\right) \left(-a^2 \ell ^2+2 \rpl^4+\rpl^2 \ell ^2\right) \left(a^4 \mathcal{L}_{\phi _1}-2 a^3 \ell ^2+4 a^2 \ell ^2 \mathcal{L}_{\phi _2}-2 a \ell ^4+3 \ell ^4
   \mathcal{L}_{\phi _1}\right)}{\rpl \left(a^2+\ell ^2\right) \left(\rpl^2+\ell ^2\right) \left(-a^4\ell ^2+a^3 \mathcal{L}_+ \left(\rpl^2+3 \ell ^2\right)-5 a^2 \ell ^2
   \left(\rpl^2+\ell ^2\right)+a \ell ^2 \mathcal{L}_+ \left(5 \rpl^2+3 \ell ^2\right)-\rpl^2 \ell ^4\right)}\cr&&\cr
&&\delta b=\frac{\tilde{\delta}\rpl \left(a^2-\ell ^2\right) \left(-a^2 \ell ^2+2 \rpl^4+\rpl^2 \ell ^2\right) \left(a^4 \mathcal{L}_{\phi _2}-2 a^3 \ell ^2+4 a^2 \ell ^2 \mathcal{L}_{\phi _1}-2 a \ell ^4+3 \ell ^4
   \mathcal{L}_{\phi _2}\right)}{\rpl \left(a^2+\ell ^2\right) \left(\rpl^2+\ell ^2\right) \left(a^4 \left(-\ell ^2\right)+a^3 \mathcal{L}_+ \left(\rpl^2+3 \ell ^2\right)-5 a^2 \ell ^2
   \left(\rpl^2+\ell ^2\right)+a \ell ^2 \mathcal{L}_+ \left(5 \rpl^2+3 \ell ^2\right)-\rpl^2 \ell ^4\right)}   \cr&&
\label{delta_ab}
\eea
Here we denote the change in the outer radius as $r_0\rightarrow r_0+\tilde{\delta}\rpl$ so as to differentiate from the $\delta\rpl$ used in the near horizon limits \eqref{strict_ext_5_MP},\eqref{limits_ads_MP_5}. 
The infinitesimal temperature $T_H$ is therefore related to the change in the outer horizon $\tilde{\delta}\rpl$ as
\be
\tilde{\delta}\rpl=2\pi\frac{\rpl^2\left(\rpl^2+\ell^2\right)}{4\rpl^2+\ell^2} T_H.
\label{delta_rp_T_H}
\ee
Therefore taking the near extremal limit of the ADM mass $M$ and entropy $\mathcal{S}$ as $\rmi=r_0-\tilde{\delta}\rpl$ \& $\rpl\rightarrow r_0+\tilde{\delta}\rpl$ reveals that the excess mass is related to $T_H$ as\footnote{One can show that $\Delta M_{\rm Casimir}=0$.}
\be
\Delta M=\frac{8\pi^2 r_0^4\ell^2(\ell^2-r_0^2)}{(\ell^2-2r_0^2)^2(4r_0^2+\ell^2)(1-\mu_0\,\Ll_+)}T_H^2,\hspace{0.5cm}\Delta\mathcal{S}=\frac{16\pi^2r_0^4\ell^4(\ell^2-r_0^2)}{(\ell^2-2r_0^2)^2(4r_0^2+\ell^2)}T_H
\label{Delta_M_S_Kerr_MP_5}
\ee
where $\Ll_+=\Llo+\Llt$. The above relations are independent of how we choose to label the change in the outer horizon as the we measure the change in $M$ \& $\cS$ terms of $T_H$.
We therefore 	 see the first law of black hole thermodynamics for small fluctuations $\delta T_H$  of the temperature as
\be
\delta M=\frac{T_H}{(1-\mu_0\, \Ll_+)}\delta\mathcal{S}
\label{first_law_5_0}
\ee
which is consistent with the first law
\bea
&&\delta M= T_H\,\delta\cS+\mu (\delta J_{\varphi_1}+\delta J_{\varphi_2}),\cr&&\cr
{\rm where}&&\delta J_{\varphi_1}=\Llo\,\delta M\,\,\&\,\,\delta J_{\varphi_2}=\Llt\,\delta M
\eea 
We would like to get the above dependence of $\Delta M$ and $\Delta \mathcal{S}$ on $T_H$- and as a consequence the first law,  from the near horizon JT description for the variation \eqref{Delta_J_Kerr_MP_5}.
\subsection{NHNext JT$_\Ll$}
\label{MP_Kerr_AdS_5_NHNext_JT_analysis}
We begin with reproducing the extremal value of the entropy $\cS_{ext}$ from the topological term $I_{(0)}$ in \eqref{JT_ads_action_Kerr_MP_5}. For this we need the only the on-shell value of the 2d metric whihc corresponds to the near extremal $AdS_2$ throat region of the Kerr geometry as in \eqref{Next_AdS_5_1}
\be
ds^2_{(2)(E)}=\frac{dR^2}{R^2-\delta\rpl^2}+(R^2-\delta\rpl^2)dT^2
\label{ads_2_ads_5_NH}
\ee
This solves the metric equation of motion $R_{(2)}+2=0$ \eqref{eom_JT}. Evaluating $I_{(0)}$ on-shell we use
$\delta\rpl=2\pi T^{(2)}_H=2\pi/\beta^{(2)}$ to denote the 2d temperature we have
\bea
I_{(0)}&=&-\frac{\pi}{G_5}\frac{\Phi_0^3}{\gamma_0}\left\lbrace\int d^2x\sqrt{g_{(2)}}\, \,R+2\int_\partial dx\sqrt{h}\,\,K\right\rbrace\cr&&\cr
&=&-\frac{\pi}{G_5}\frac{\Phi_0^3}{\gamma_0}\left[\int_{\delta\rpl}^{R_\infty}\hspace{-0.3cm}\int_0^{\beta^{(2)}}\hspace{-0.5cm}d^2x\sqrt{g}R+2\int_0^{\beta^{(2)}}\hspace{-0.5cm}dT\,\,\sqrt{h}\,K\right]\cr&&\cr
&=&-\frac{\pi}{G_5}\frac{\Phi_0^3}{\gamma_0}2\delta\rpl\beta^{(2)}=\frac{4\pi^2\Phi^3_0}{G_{(5)}\gamma_0}=\cS_{ext}
\label{I_0_S_ext_check_Kerr_MP_5}
\eea
where we used the horizon volume \eqref{transverse_volume_Kerr_MP_5} in terms of $\Phi_0$. As in the previous case of the Kerr black hole in $AdS_4$ we see that the  only information we need for this check to work out is how the 2d Newton's constant($i.e.$ constant in front of the integral for $I_{(0)}$) is related to the extremal horizon volume.
\\\\
We next turn to determining $\Delta\cS$ and $\Delta M$. For this we need 2 further geometric data relating the near horizon geometry to the higher dimensional one. As argued below \eqref{metric_kruskal_L_Kerr_MP_5} the
geometric argument relates the near horizon $AdS_2$ temperature $T^{(2)}_H$ to $T_H$ as
\be
T_H^{(2)}=\frac{T_H}{1-\mu\,\Ll_+}=\frac{\delta\rpl}{2\pi}
\label{T_2_T_H_delta_rp_5_MP}
\ee
We next need on-shell stationary value of the dilaton $\phi$ at the $AdS_2$ horizon $R=\delta\rpl$ to get the fluctuation of the horizon volume close to extremality, this is
\be
\phi=R\phi_\partial\implies\phi\vert_{\delta\rpl}=\delta\rpl\phi_\partial
\label{static_dilaton_ads_2_Kerr_MP_5}
\ee
Given that the JT action was derived relating the horizon volume $V_H$ to the fluctuation $\phi$ of the dilaton as
\be
V_H=\frac{16\pi^2\Phi_0^3}{\gamma_0}\left(1+3 \,\phi\vert_{\delta\rpl}\right)
\label{trasverse_volume_ads_2_Kerr_MP_5}
\ee
we find that $\phi_\partial$ is given by 
\be
\phi_\partial=\frac{1}{\delta\rpl}\frac{\Delta \Phi}{\Phi_0}
\label{phi_partial_Kerr_MP_5}
\ee
The change $\Delta \Phi$ is computed as the change in the transverse volume of the horizon $V_H$ \eqref{transverse_volume_Kerr_MP_5} in the full metric  \eqref{Kruskal_coordinates_equator_Kerr_MP_5} as we take  the near horizon limit  $\lambda\rightarrow 0$ scaling 
\be
\rmi=r_0-\lambda \,\cA\, \delta\rpl,\hspace{0.3cm}\rpl=r_0+\lambda \,\cA\, \delta\rpl
\ee
to first order in $\lambda$ with $\cA$ as in \eqref{strict_ext_5_MP}. $\Phi_0$ is given in \eqref{Next_AdS_5_1} corresponding to the extremal value of $V_H$. We thus find
\be
\frac{\Delta\Phi}{\Phi_0}=\frac{r_0(\ell^2-r_0^2)}{3(4r_0^2+\ell^2)}(1-\mu\,\Ll_+)\delta\rpl
\label{Delta_Phi_by_Phi_Kerr_MP_5}
\ee
The entropy change is then simply given by change in extremal volume $V_H$, thus
\be
\frac{\Delta\cS}{\cS_{ext}}=3\left.\phi\right\vert_{R=\delta\rpl}=\delta\rpl \phi_\partial
\label{Delta_S_by_S_ads_2_Kerr_MP_5}
\ee
where we use the expressions \eqref{T_2_T_H_delta_rp_5_MP},\eqref{phi_partial_Kerr_MP_5}. Using \eqref{Delta_Phi_by_Phi_Kerr_MP_5} the above $rhs$ matches the value of $\Delta\cS/\cS_{ext}$ computed using \eqref{Delta_M_S_Kerr_MP_5} and \eqref{T_H_S_Kerr_MP_5} thus reproducing the excess entropy over extremality.
\\\\
The change  over the extremal mass is obtained by simply computing the 2d $ADM$  mass as before
\be
M^{(2)}=\frac{\delta\rpl}{16\pi G_{(2)}}\phi|_{R=\delta\rpl}=\frac{\delta\rpl^2}{16\pi G_{(2)}}\phi_\partial,\hspace{1cm}{\rm where}\hspace{0.3cm}G_{(2)}=\frac{\gamma_0 G_{(5)}}{48\pi^2 \Phi_0^3}
\label{M_ads_2_5_MP}
\ee
Substituting the value of $\delta\rpl$ and $\phi_\partial$ from  \eqref{T_2_T_H_delta_rp_5_MP} and \eqref{phi_partial_Kerr_MP_5} respectively we find that it exactly reproduces the expression for $\Delta M$ in \eqref{Delta_M_S_Kerr_MP_5} as
\be
M^{(2)}=\frac{T_H^{(2)}}{2}\Delta\cS=\Delta M
\label{M_2_JT_Kerr_MP_5}
\ee
Therefore the JT$_\Ll$ theory correctly reproduces the change  $\Delta M$ \& $\Delta\cS$ as a function of the near extremal temperature $T_H$ as in \eqref{Delta_M_S_Kerr_MP_5}. 
Consequently for infinitesimal variations in $\delta T_H$ we find that the first law \eqref{first_law_5_0} is also reproduced about extremality.
\section{Discussions \& Conclusions}
\label{D&C}
In this paper we have argued that the thermodynamics of near extremal Kerr geometries for perturbations both to its mass and angular momentum are captured by the Jackiw-Teitelboim model. 
We considered change in the extremal mass and angular momenta such that $\Delta J=\Ll\,\Delta M$ for Kerr $AdS_4$ and $\Delta J_{\varphi_1,\varphi_2}=\Ll_{\varphi_1,\varphi_2}\,\Delta M$ for Myers-Perry Kerr black hole in $AdS_5$ such that the resultant geometries are near extremal. Therefore we focus on the thermodynamics of near-extremal Kerr black holes in the canonical ensemble where a linear combination of ADM charges are held fixed.
We do this by explicitly generalizing the near horizon limits prevalent in literature by approaching the outer horizon along in-going null rotating geodesics parametrized by specific angular momenta $\Ll$ for Kerr $AdS_4$ and $\Ll_{\varphi_1,\varphi_2}$ for MP Kerr $AdS_5$. We discover distinct IR geometries parametrized by the specific angular momenta used to obtain the near horizon metric. These geometries- labelled by $\Ll$ for $next$ Kerr $AdS_4$ and $\Ll_\pm$ for $next$ MP Kerr $AdS_5$, cannot be related to each other by simple change of their near horizon coordinates $c.f.$ subsections- \ref{L_0_to_L_neq_0} \& \ref{L_pm_0_to_L_pm_neq_0}. The near horizon JT action obtained for such IR geometries- referred to as JT$_\Ll$, differs in the manner in which it is embedded in the IR of theory dual to the higher dimensional $next$ geometry.
We explicitly derive the small temperature dependence of excess mass and entropy for such black holes from the JT$_\Ll$ model in the near horizon region obtained using such limits. Setting the specific angular momenta to zero from the start amounts to reproducing similar analysis prevalent in literature \cite{Nayak:2018qej,Moitra:2019bub}. 
\\\\
A distinctive feature of  the JT$_\Ll$ analysis is that the near horizon throat region- described by an $AdS_2$ geometry, has a temperature given by $T_H^{(2)}=T_H/(1-\mu_0\,\Ll)$ as seen in \eqref{T_2_T_H_Kerr_4} \&\eqref{T_H_T_H_Kerr_MP_5}. 
This relation is crucial for obtaining the match between the small temperature behaviour of the JT$_\Ll$ model and higher dimensional near extremal Kerr geometry. 
Given that the low energy (or large time $\gg\beta$) behaviour of the OTOCs in the dual CFT$_{4,5}$ is obtainable from the near horizon $nAdS_2/nCFT_1$ of the JT model, the analysis in this paper suggests that the Lyapunov exponent related to  perturbations to both the mass and angular momentum of  Kerr geometries is $2\pi T_H/(1-\mu_0\,\Ll) $. This is similar to the results obtained in \cite{Malvimat:2022fhd,Malvimat:2022oue} where $\lambda_L$ for late time rotating perturbations to mutual information contained in the TFD state dual to the Kerr geometries were analysed. 
In the presence of a probe matter field $\Phi_V$ the black hole geometry has its mass perturbed by $\Delta M$ dictated by the conformal dimension $\Delta_V$ of the dual operator $V$ in the CFT while the change in its angular momentum $\Delta J$ depends on the spin $s$ of $\Phi_V$ and $V$. A clear relation between the black hole parameters and those of the CFT operators is furnished in $AdS_3/CFT_2$ where heavy operators in the dual $CFT_2$ correspond to a bulk state identical to a BTZ geometry at leading order\footnote{This is true to leading order in $h/c$ \& $\bar{h}/c$, sub-leading corrections do reveal the difference between a thermal background dual to a BTZ and a background due to the presence of 2 heavy operators.  } \cite{Fitzpatrick:2015zha} with $J=\frac{12 (h-\bar{h})}{c}$ and $M=\frac{12(h+\bar{h})}{c}+2$, with $c=\frac{3\ell}{2G_{3}}$ and $\Delta=h+\bar{h}$, $s=h-\bar{h}$. Here the bulk geometry is the result of 2 identical heavy operators with $CFT_2$ weights $\{h,\bar{h}\}$ inserted at $\pm\infty$. Therefore given the near horizon analysis presented here in conjunction with the observations in \cite{Malvimat:2022fhd,Malvimat:2022oue} we posit that the OTOC of operators with spin in CFTs dual to Kerr geometry exhibit an instantaneous  Lyapunov index given by $2\pi T_H/(1-\mu\,\Ll)$ where $\Ll$ is related to the spin of the operators involved. The precise nature of such correlators would be a work we would like to pursue in the future.
\\\\
The Kerr/CFT correspondence attempts to holographically understand the IR description for rotating extremal black holes \cite{Hartman:2008pb,Lu:2008jk,Guica:2008mu}\footnote{see \cite{Compere:2012jk} for review}. Here too one first obtains the near horizon geometry in an (near)extremal setting, however it seems that- like $AdS_2$, the near horizon geometries cannot seem to contain any finite energy states \cite{Amsel:2009ev,Dias:2009ex,Hajian:2014twa}. 
There seems to be technical difficulties in writing down an effective action of gravitational interactions (like \eqref{JT_Schwarzian}) in the Kerr/CFT setting.
However, in 2d, one can systematically analyse the perturbations to the near horizon structure at (near)extremality  and account for subtleties not captured in the JT model \cite{Castro:2021fhc,Castro:2021csm,Castro:2019crn,Castro:2018ffi}. Here, although the deep IR thermodynamic response is captured by the JT model \cite{Moitra:2018jqs}, there are important corrections in the IR from the higher Kaluza-Klein modes and from sub-leading terms in the near horizon limit. 
For $ex.$ the authors in \cite{Castro:2018ffi,Castro:2021csm} show that beside the JT theory the effective gravitational theory has other fields relevant for understanding the IR of $AdS_5/CFT_3$ holographic system especially when coupled to probe matter. Each of the above analyses requires working with a near horizon limit for Kerr black holes and the results of this paper suggest an important peculiarity as the near horizon limits in these analyses are obtained for $\Ll=0$ ($\Ll_\pm=0$). As the near horizon geometry for $\Ll\neq 0$ ($\Ll_\pm\neq 0$) is \emph{not} simply related to the $\Ll=0$ ($\Ll_\pm=0$) case by a change of near horizon coordinates, the IR dynamics captured by the JT$_\Ll$ model requires non-trivial or \emph{relevant} corrections to the JT model obtained with $\Ll=0$ ($\Ll_\pm=0$) from the higher dimensional geometry\footnote{\emph{Note,} the Newman-Penrose formalism-required for such analyses, can also be cast in terms of rotating null in-out going geodesics as the formalism is based on veilbeins. }.
\\\\
Many body quantum chaos in holographic theories dual to Kerr geometries have also recently been under investigations where Teukolsky formalism is employed for the studying the energy fluctuations in the boundary theory to in-falling perturbations \cite{Blake:2021hjj,Amano:2022mlu} . The energy-energy correlators are expected to exhibit a phenomena of \emph{pole-skipping} \cite{Blake:2017ris,Blake:2018leo,Blake:2021wqj} where the poles skipped in the complex frequency plane correspond to the value of $\lambda_L$. The analysis presented here predicts that these analyses when generalized for rotating in-falling shockwaves must see a $\lambda_L$ as in \eqref{Lyapunov_MI_Kerr_4} $i.e.$ as seen in \cite{Malvimat:2022fhd,Malvimat:2022oue}. It would be interesting to see how the butterfly velocities $v_B$ are modified due to $\Ll$.
\\\\
The recent results of the gap in the density of states  for near BPS non-supersymmetric \cite{Heydeman:2020hhw} and  supersymmetric \cite{Boruch:2022tno}  configurations analysed by the JT model can also be reanalysed in this light $i.e.$ in a canonical ensemble where angular momentum is allowed to change linearly with mass. Given the analysis presented here one can expect the large time effects as seen in \cite{Lin:2022rzw,Lin:2022zxd} to change appropriately.
\section*{Acknowledgements} The author is grateful for comments by  Joan Simon on the manuscript and discussions with Daniel Grumiller. This work is  supported by the Austrian science fund FWF $via$ the Lise Meitner grant M2882-N.
\appendix
\section{Conical deficit}
\label{conical_deficit}
Here we show that the Kerr $AdS_4$ metric written in the form \eqref{Kerr_Kruskal_4} in terms of in-out going null geodesics sees a conical deficit given by \eqref{kappa_Kerr_4} $i.e.$ 
\be
\kappa=\frac{2\pi T_H}{1-\mu\cL}
\ee 
This conical deficit is measured about $\tau$ instead on the Boyer-Lindquist time $t$ where
\be
t=\frac{\tau+\Ll\, z}{1-\mu \,\mathcal{L}},\hspace{0.3cm}\varphi=\frac{(1-a\mathcal{L}/\l^2)\,z+\Omega_{\varphi}\tau}{1-\mu \,\mathcal{L}}.
\label{t_phi_to_tau_z_Kerr_4_appendix}
\ee
The smooth Kruskal coordinates used to write the metric along the rotating in-out going null geodesics renders the line element as ($c.f.$ \eqref{Kruskal_coordinates_equator_Kerr_4}\eqref{LC_coordinates_Kerr_4})
\bea
&&ds^2=\frac{F}{\kappa^2 UV}dUdV + h\,(dz+h_\tau d\tau)^2 +g\, (d\theta+g_\tau d\tau)^2,\hspace{0.4cm} g_\tau(\pi/2)=0 \cr&&\cr
{\rm where}&&d\tau=\frac{1}{2UV}(UdV-VdU)-\tilde{g}'d\theta
\label{Kerr_Kruskal_4_appendix}
\eea
As we would like to measure the conical deficit about $\tau$ we rewrite the metric as
\be
ds^2=F\left(\frac{dr^2}{f^2}-d\tau^2\right) + h\,(dz+h_\tau d\tau)^2 +g\, (d\theta+g_\tau d\tau)^2
\label{rotating_null_appendix}
\ee
where $f$ can be inferred from the expression for $dr_*$ in \eqref{LC_coord_Kerr_4}. We next note that $\kappa$ is obtained be demanding smoothness of the Kruskal coordinates at $\rpl$ and its form is
\bea
&&\kappa=\frac{1}{2}\xi^r_+\cdot\partial_r F\vert_{\rpl}, \hspace{0.2cm}{\rm where}\,\,\xi_rdr=\frac{dr}{f}, \& \,\,\xi^r=\frac{f}{F}\cr&&\cr
\implies && \kappa=\frac{1}{2}f\partial_r \log F\vert_{\rpl},\hspace{0.3cm}i.e.\hspace{0.2cm} F'\vert_{\rpl}=\left.\frac{2\kappa F}{f}\right\vert_{\rpl}
\label{kappa_F_appendix}
\eea
This can be used while expanding $F$ around $\rpl$ as
\be
F=(r-\rpl)F'(\rpl)+\frac{(r-\rpl)^2}{2}F''(\rpl)+\dots
\label{F_expand_appendix}
\ee
The above relations \eqref{F_expand_appendix},\eqref{kappa_F_appendix} can be used to expand the $dr$ component around $\rpl$ as
\be 
\frac{dr}{f}=\left(\frac{F}{f}\right)_{\!\!\rpl}\!\!\frac{dr}{F}\sim\frac{dr}{(r-\rpl)2\kappa}=\frac{dR}{2\kappa R},{\rm where}\,\,\, R=r-\rpl
\ee
Therefore the $\{r,\tau\}$ components of the line element of the above metric \eqref{rotating_null_appendix} can be expanded around $\rpl$ as
\be
F\left(\frac{dr^2}{f^2}-d\tau^2\right)=F\left(\frac{dR^2}{4\kappa^2 R^2}-d\tau^2\right)=\frac{F}{4\kappa^2 R^2}\left(dR^2-R^2(2\kappa)^2d\tau^2\right)
\ee
One can easily see that if the above metric were Euclideanised then the conical deficit about the Eulcideanised $\tau$ is equal to $\kappa$.  
This also becomes obvious if one  compares  with a near horizon expansion of  a 2d $AdS_2$ geometry with a horizon at $\rpl$
\be
\left(\frac{dr^2}{(r^2-\rpl^2)}-(r^2-\rpl^2)d\tau^2\right)\overset{r\rightarrow\rpl}{\longrightarrow}\frac{1}{R(R+2\rpl)}\left(dR^2-R^2(2\rpl^2)^2d\tau^2+\dots\right)
\ee
where again we set $R=r-\rpl$ and the $\dots$ indicate sub-leading terms in about $R=0.$ 
\section{JT Thermodynamics}
\label{JT_appendix}
In this appendix we review the computation of the Free energy and the ADM mass $M^{(2)}$ for the JT action. We assume that the 2d Euclidean metric is that of a black hole in $AdS_2$ with a temperature $T^{(2)}_H=\frac{\delta\rpl}{2\pi}=\left(\beta^{(2)}\right)^{-1}$
\be
ds^2_{(2d)(E)}=\frac{dR^2}{R^2-\delta\rpl^2}+(R^2-\delta\rpl^2)dT^2
\label{2d_metric_appendix}
\ee 
Given this the topological term $I_{(0)}$ in the action \eqref{JT_action} evaluates to
\bea
I_{(0)}&=&-\frac{1}{16\pi \alpha G_{(2)}}\left[\int_{\delta\rpl}^{R_\infty}\hspace{-0.3cm}\int_0^{\beta^{(2)}}\hspace{-0.5cm}d^2x\sqrt{g}\,R_{(2)}+2\int_0^{\beta^{(2)}}\hspace{-0.5cm}dT\,\,\sqrt{h}\,K_{(2)}\right]\cr&&\cr
&=&-\frac{1}{16\pi \alpha G_{(2)}}2\delta\rpl\beta^{(2)}=-\cS_{ext}
\label{S_ext_appendix}
\eea
where contributions at the boundary $R_\infty$ cancel from the bulk and boundary parts. the fact that the above action evaluates to $\cS_{ext}$ depends on how $G_{(2)}$ and $\alpha$ are related to the extremal value of the horizon volume which needs to be specified.
\\\\
The ADM mass $M^{(2)}$ can be computed using 2 methods using the renormalized JT action.
\be
I_{(JT)}=-\frac{1}{16\pi G_{(2)}}\left[\int\!d^2x\,\sqrt{g_{(2)}}\,\phi\left(R_{(2)}+2\right)+2\int_\partial\!dx\,\sqrt{h_{(2)}}\,\phi\,\left(K_{(2)}-1\right)\right]
\label{JT_renorm_appendix}
\ee
The bulk integral vanishes on-shell and we only need the boundary integral above.
The first is using the renormalized Brown-York tensor. Since the boundary is one dimensional the renormalized Brown-York stress tensor must simply evaluate to $M^{(2)}$
\be 
M^{(2)}:=T_{ab}=\left.\frac{2}{\sqrt{\gamma}}\frac{\delta I_{JT}}{\delta \gamma^{ab}}\right\vert_{\gamma=\gamma_0}
\label{Brown_York_stress_tensor_appendix}
\ee
where we take the boundary metric to be $\gamma_{ab}dx^adx^b=\gamma^2(t)dT^2$, after having functionally differentiated $I_{JT}$ $w.r.t.$ $\gamma^{ab}$ we put set it equal to the boundary metric $\gamma_0^2dT^2$ were $\gamma_0=1$. The $AdS_2$ bulk metric with arbitrary boundary metric is
\be
ds^2_{2d}=\frac{dR^2}{R^2-\delta\rpl^2}+\gamma^2(t)(R^2-\delta\rpl^2)dT^2
\label{JT_ads_2_arbit_boundary_metric_appendix}
\ee
which solves the $eom$ $R_{(2)}+2=0$. The stationary dilaton solving the eom \eqref{eom_JT} for the above metric again has the same form as in \eqref{dilaton_sol_gen_JT} $i.e.$
\be
\phi=R\,\phi_\partial
\label{static_dilaton_sol_appendix}
\ee
Using the metric \eqref{JT_ads_2_arbit_boundary_metric_appendix} and the above value of the dilaton and setting $\gamma(t)=1$ after evaluating \eqref{Brown_York_stress_tensor_appendix} we find
\be
M^{(2)}=\frac{\delta\rpl^2\phi_\partial}{16\pi G_{(2)}}=\Delta M
\label{JT_M_appendix}
\ee
The second method is to evaluate the on-shell value of $I_{JT}$ for the metric \eqref{2d_metric_appendix} with a stationary solution of the dilaton $\phi= R\,\phi_\partial$ and using the thermodynamic relation
\be
I_{JT}^{(OS)}=\beta^{(2)}F=\beta^{(2)}M^{(2)}-\Delta\cS
\label{JT_Gibbs_Free_Energy_appendix}
\ee
where $\Delta\cS$ is the excess entropy over $\cS_{ext}$. 
\be
I^{(OS)}_{JT}=-\frac{1}{16\pi G_{(2)}}\beta^{(2)}\delta\rpl^2\,\phi_\partial
\label{JT_on_shell_appendix}
\ee
where the boundary integral is from $[0,\beta^{(2)}]$. $\Delta\cS$ is given by value of the dilaton at the $AdS_2$ horizon as it measures the fluctuation of the horizon volume about extremality
\be
\Delta\cS=\left.\alpha\,\cS_{ext}\,\phi\right\vert_{R=\delta\rpl}=\alpha\,\cS_{ext}\,\delta\rpl\,\phi_\partial
\label{Delta_S_appendix}
\ee
Using \eqref{S_ext_appendix} with the above expression for $\Delta\cS$ in \eqref{JT_Gibbs_Free_Energy_appendix}  we get a match with \eqref{JT_M_appendix}.

\bibliographystyle{JHEP.bst}
\bibliography{bulk_syk_soft_modes.bib}
\end{document}